\documentclass[11pt,a4paper]{article}
\usepackage{amssymb}
\usepackage{amsmath}
\usepackage{jheppub}

\newcommand{\be}{\begin{equation}}
\newcommand{\ee}{\end{equation}}
\newcommand{\ba}{\begin{align}}
\newcommand{\ea}{\end{align}}

\title{Cosmologies in Horndeski's second-order \\ vector-tensor theory}

\author[a]{John D. Barrow,}
\author[b]{Mikjel Thorsrud}
\author[a,b]{and Kei Yamamoto}

\affiliation[a]{DAMTP, University of Cambridge, \\
Wilberforce Road, Cambridge, CB3 0WA, United Kingdom}
\affiliation[b]{Institute of Theoretical Astrophysics, University of Oslo, \\
P.O. Box 1029 Blindern, N-0315 Oslo, Norway}
\emailAdd{J.D.Barrow@damtp.cam.ac.uk}
\emailAdd{mikjel.thorsrud@astro.uio.no}
\emailAdd{K.Yamamoto@damtp.cam.ac.uk}
\abstract{Horndeski derived a most general vector-tensor theory in which the vector field respects the gauge symmetry and the resulting dynamical equations are of second order.
The action contains only one free parameter, $\lambda$, that determines the strength of the non-minimal
coupling between the gauge field and gravity. We investigate the cosmological consequences of this action and discuss observational constraints. For $\lambda<0$ we identify singularities where the deceleration parameter diverges within a finite proper time. This effectively rules out any sensible cosmological application of the theory for a negative non-minimal coupling.  We also find
a range of parameter that gives a viable cosmology and study the phenomenology for this case. Observational constraints on the value of the coupling are rather weak since the interaction is
higher-order in space-time curvature.}
\begin{document}

\maketitle

\section{Introduction}

In the past few decades, modifying the Einstein's theory of gravitation has
been an active area of research \cite{Clifton2011}, driven chiefly by the search
for different varieties of inflation, the desire of some to explain flat
galaxy rotation curves without dark matter \cite{Milgrom2008}, and the challenge of
explaining why the expansion of the universe started to accelerate at late
times. It is also natural to question the validity of general relativity,
not least because of its ultra-violet behaviour which does not give a
well-defined quantum field theory. There is a growing prospect of testing
any such deviations from general relativity in very strong gravity fields by
searching for the signatures of gravitational waves created by high-energy
astrophysical phenomena, such as black hole mergers, or by scrutinising
detailed observations of the microwave background anisotropy and statistics
in the light of particular theories of inflation. While there are a plethora
of inflationary models, it is difficult to modify general relativity without
spoiling its appealing features and typical modifications end up introducing
new scalar degrees of freedom, as was the case in the case of Brans-Dicke
gravity \cite{Brans1961} and its scalar-tensor generalisations \cite{Nordtvedt1970, Wagoner1970, Barrow1990}, or $f(R)$
lagrangian theories of gravity \cite{Ruzmaikina1971, Barrow1983, Barrow1988, Sotiriou2010, Sotiriou2011}. 

Maxwell's classical theory of electromagnetism has also been extremely well
tested and there are already strict constraints on potential modifications
such as a non-zero photon mass \cite{GOLDHABER1971, Barnes1979, Barrow1984, Dolgov1981} or varying fine structure
constant \cite{Webb1999, Murphy2008, Bekenstein1982, Sandvik2002, Barrow2002, Barrow2002a, Barrow2002b, Uzan2003, Barrow2012}. However, there has been renewed interest in modified
electromagnetism in cosmology and there have been attempts to incorporate
its effects into the dynamics of early universe, particularly during
inflation, or to provide an explanation for cosmological magnetic fields 
\cite{Turner1988, Ratra1992, Calzetta1998, Giovannini2000, Lambiase2004, Kunze2005, Bamba2007, Kunze2008, Campanelli2008, Campanelli2008a, Bamba2008a, Bamba2008b, Campanelli2009, MosqueraCuesta2009, Campanelli2009a, Kunze2010}. Simple vector fields themselves are known to have difficulties in producing inflation. The time variation of the vector field
is governed by the covariant time derivative of the field. Since the
Christoffel symbols for an expanding cosmological model are of order the
Hubble expansion rate it is not possible for the vector field to satisfy a
slow-roll condition in the way that a scalar field can \cite{Ford1989}. However
Einstein-aether theories offer an alternative that permits inflation \cite%
{Donnelly2010, Gasperini1985, Carroll2004, Lim2005, Li2008, Zuntz2008, Armendariz-Picon2010, Zlosnik2008, Meng2011, Nakashima2010, Barrow2012, Clifton2011}. In addition, although it was noted that a non-minimal
coupling to the space-time curvature could drive accelerated expansion \cite%
{Koivisto2008a, Golovnev2008, Bamba2008,jimenez08}, these scenarios suffer from
various instabilities created by additional degrees of freedom arising from
the lost gauge symmetry \cite{himmetoglu08,carroll09,himmetoglu08b,Koivisto2009, Himmetoglu2009, Golovnev2009}. A
different type of extension of the Maxwell case is provided by the extension
to Yang-Mills fields where there can be chaotic behaviour and arbitrarily low
levels of anisotropy \cite{Barrow1998, Jin2005, Barrow2005}.

Recently, it has been mentioned that a vector-tensor theory, first proposed
by Horndeski in 1976 \cite{horndeski76}, could lead to an instability of
conventional inflationary universe through the non-minimal coupling between
the vector field and gravity \cite{farese09}. The action was derived by
demanding second-order dynamical equations that reduce to Maxwell's equations when evaluated on a Minkowski background and conservation of the $U(1)$ current. These requirements
result in only one additional term in the Lagrangian, and therefore a single
free coupling parameter. Later, it was noticed that this theory falls into a
special class of Kaluza-Klein reductions from higher-dimensional Lovelock
invariants \cite{buchdahl79, Lovelock1971, Lovelock1972}. In contrast to the Horndeski scalar
field theory \cite{horndeski74}, which has been discussed in attempts to
construct the most general viable scalar-tensor theory recently, \cite%
{scalartensor1,scalartensor2,scalartensor3,scalartensor4,scalartensor5,scalartensor6}%
, except for a brief examination of the static electromagnetism arising from
this action \cite{horndeski77}, it appears to have escaped attention. Apart
from being briefly mentioned in \cite{farese09}, its cosmological
consequences have not been studied. 

In this paper, we will investigate the simplest cosmological model, which is
well understood in the minimally coupled case of a Maxwell electromagnetic
field \cite{blanc97, collins72}, containing a perfect fluid and a vector
field whose dynamics are described by the Horndeski Lagrangian. We find the
following results;

\begin{enumerate}
\item The instability found in \cite{farese09} for negative values of the
coupling constant persists in the nonlinear regime and the universe
eventually hits a singularity;

\item For a positive coupling constant, the electric field can still be
amplified during the radiation-dominated era while giving a viable cosmology
subject to some constraints on the allowed expansion rate changes at the
epoch of primordial nucleosynthesis.
\end{enumerate}

The first result effectively rules out any interesting cosmological
application of this theory with a negative coupling. For a positive coupling
constant, the modification is rather tame and an enormous value of the
coupling in units of the Planck mass is allowed because of the higher-order
nature of the modified term. However, the dynamics is of a phenomenological
interest.

The article is organised as follows. In the next section, the theory is
introduced and the modified Einstein-Maxwell equations are presented.
Section \ref{ch:electric} is the main part of the article where the dynamics of purely
electric component are studied in an axisymmetric Bianchi type I universe.
In section  \ref{ch:magnetic}, we repeat the previous analysis for magnetic component.
Section  \ref{ch:constraints} discusses observational constraints and in section  \ref{ch:conclusion} we summarise
our principal results.

\section{Horndeski's second-order vector-tensor theory}

In 1976, Horndeski showed that the general Lagrangian that can be
constructed from a metric $g_{ab}$ and a vector field $A_{a}$ in
four-dimensional space-time that satisfies the following conditions \cite{horndeski76}:

\begin{enumerate}
\item the field equations contain at most second-order derivatives of $%
g_{ab} $ and $A_{a}$ (and do contain a second-order term);

\item the dynamical equations for $A_{a}$ respect charge conservation i.e. $%
\nabla _a (\partial \mathcal{L} / \partial A_a ) = 0$;

\item the dynamical equations for $A_{a}$ reduce to Maxwell's equations when
evaluated on Minkowski space-time;
\end{enumerate}

takes the following form: 
\begin{equation}
\mathcal{L}=\frac{M_{pl}^{2}}{2}\sqrt{-g}R-\frac{1}{4}\sqrt{-g}F_{ab}F^{ab}+%
\mathcal{L}_{H},
\label{eq:fulllagrangian}
\end{equation}%
where $M_{pl}$ is the reduced Planck mass, $R$ is the Ricci scalar and $F_{ab}=\partial_a A_b - \partial_b A_a$ is the Faraday tensor.
The last term is Horndeski's modification which can be expressed in several different ways as
\begin{eqnarray}
\mathcal{L}_{H} &=&-\frac{3\lambda }{2M_{pl}^{2}}\sqrt{-g}\delta _{\
efkl}^{abcd}F_{ab}F^{ef}R_{cd}^{\ \ kl} \\
&=&\frac{\lambda }{4M_{pl}^{2}}\sqrt{-g}F_{ab}F^{cd}\ast R\ast _{\ \ cd}^{ab}
\\
&=&-\frac{\lambda }{4M_{pl}^{2}}\sqrt{-g}\left(
RF_{ab}F^{ab}-4R_{ab}F^{ac}F_{\ c}^{b}+R_{abcd}F^{ab}F^{cd}\right) .
\end{eqnarray}%
The dimensionless non-minimal coupling constant $\lambda $ is the only
parameter of the theory.  Our aim is to investigate the cosmological consequences with an arbitrary value of $\lambda$ and to determine the parameter range yielding viable phenomenology. The other terms in the Lagrangian (\ref{eq:fulllagrangian}) are normalized so that it reduces to
the Einstein-Maxwell theory when $\lambda = 0$. Using the Levi-Civita tensor $%
\eta _{abcd}$, we defined the generalised Kronecker's delta by 
\begin{equation*}
\delta _{\ efkl}^{abcd}=\delta _{\ [e}^{a}\delta _{\ f}^{b}\delta _{\
k}^{c}\delta _{\ l]}^{d}=-\frac{1}{24}\eta ^{abcd}\eta _{efkl} 
\end{equation*}%
and the double dual of Riemann by 
\begin{equation*}
\ast R\ast _{\ \ cd}^{ab}=\frac{1}{4}\eta ^{abef}\eta _{cdkl}R_{ef}^{\ \
kl}. 
\end{equation*}%
$\mathcal{L}_{H}$ was later identified with the Lagrangian obtained by
Kaluza-Klein reduction from the five-dimensional Lovelock invariant 
\begin{equation*}
K=\tilde{R}_{abcd}\tilde{R}^{abcd}-4\tilde{R}_{ab}\tilde{R}^{ab}+\tilde{R}%
^{2} 
\end{equation*}%
where $\tilde{R}_{abcd}$ is the Riemann tensor in five dimensions \cite%
{buchdahl79}. 

In this paper we use the sign conventions of \cite{MTW} for the metric,
Ricci and Riemann tensors which are different from those adopted in the
previous studies of this model \cite{horndeski76,horndeski77}. The dynamical equations derived from this
Lagrangian are given as follows:

\begin{description}
\item[ Variation with respect to $g_{ij}$] 
\begin{equation}
M_{pl}^{2}G_{ij}=F_{ia}F_{j}^{\ a}-\frac{1}{4}F_{ab}F^{ab}g_{ij}+\tau _{ij},
\label{eq:tensor}
\end{equation}%
where 
\begin{equation}
\tau _{ij}=\frac{\lambda }{M_{pl}^{2}}\left( \nabla_a \ast F_{ib}\nabla
^{b}\ast F_{\ j}^{a}+F^{ab}F_{\ a}^{c}\ast R\ast _{ibjc}\right) ,
\end{equation}

\item[Variation with respect to $A_i $] 
\begin{equation}
\nabla _a F^{ia} -\frac{\lambda }{M_{pl}^2} \nabla _a F_{bc} \ast R\ast
^{iabc} =0 .  \label{eq:vector}
\end{equation}
\end{description}

We define the dual Faraday tensor as usual:%
\begin{equation*}
\ast F_{ab}=\frac{1}{2}\eta _{abcd}F^{cd}. 
\end{equation*}
In ref.\cite{farese09}, it was observed that (\ref{eq:vector}) evaluated on
a Friedmann-Lema\^{\i}tre-Robertson-Walker (FLRW) background could lead to
an instability of the vector field. Ignoring the spatial gradient term, the
solution for the comoving electric field strength $E$ evaluated on this
background is given by 
\begin{equation}
E=\frac{E_{0}}{a^{2}}\left( 1+2\lambda \frac{H^{2}}{M_{pl}^{2}}\right) ^{-1},
\label{eq:esol1}
\end{equation}%
where $E_{0}$ is an integration constant, $a$ is the scale factor and $H$ is
the Hubble expansion rate. When $\lambda $ is negative and $-2\lambda
H^{2}\gtrsim M_{pl}^{2}$, the energy density of the electric field can
rapidly increase and eventually diverge, even when the expansion of the
universe is accelerated. Our first goal is to take into account the back
reaction of this growing vector field and examine the fate of the
inflationary universe.

\section{Dynamics of electric fields in axisymmetric Bianchi type I universes%
}
\label{ch:electric}

In this section and the next, we set $M_{pl}^2 =1$.

\subsection{Electric fields in an anisotropic universe}
\label{chsub:electric}

In order to answer the question of back reaction, one needs to look at a
fully non-linear system and solve both (\ref{eq:tensor}) and (\ref%
{eq:vector}). The simplest generalisation of the FLRW universe that can
accommodate a vector field is the axisymmetric Bianchi type I metric given
by 
\begin{equation}
ds^{2}=-dt^{2}+e^{2\alpha (t)}\left[ e^{-4\beta (t)}dx^{2}+e^{2\beta
(t)}\left( dy^{2}+dz^{2}\right) \right] .  \label{eq:metric}
\end{equation}%
This metric is spatially flat. Note that the (mean) Hubble and shear
expansion rates are given by 
\begin{equation*}
H=\dot{\alpha},\ \ \ \ \ \sigma =\dot{\beta}. 
\end{equation*}%
We consider a homogeneous electric field along the $x$-direction, which in the gauge $A_0=0$ corresponds to the following coordinate basis components for the vector potential:   
\begin{equation*}
A_{\mu }=\left( 0,A(t),0,0\right).  
\end{equation*}%
The electric field strength seen by an observer moving with four-velocity $%
u^{\mu }=(1,0,0,0)$ is given by 
\begin{equation*}
E(t)=-\dot{A}e^{-\alpha +2\beta}, 
\end{equation*}%
where dots denote derivatives with respect to the comoving proper time $t$. 
We also include a perfect fluid with the equation of state $p=(\gamma
-1)\rho $ and constant $\gamma $. Hence, (\ref{eq:tensor}) yields the
following: 
\begin{eqnarray}
H^{2}-\sigma ^{2} &=&\lambda \left( H+\sigma \right) ^{2}E^{2}+\frac{1}{6}%
E^{2}+\frac{1}{3}\rho ,  \label{eq:Friedmann} \\
\dot{H}+H^{2} &=&-2\sigma ^{2}-\frac{1}{6}E^{2}-\frac{1}{6}(3\gamma -2)\rho -%
\frac{\lambda }{6}\left( H+\sigma \right) \left( H+7\sigma \right) E^{2}
\label{eq:raychaudhuri} \\
&&+\frac{\lambda E^{2}}{6}\left[ \lambda \left( H+\sigma \right) ^{2}E^{2}+%
\frac{1}{2}E^{2}-(\gamma -1)\rho +4\left( H+\sigma \right) \frac{\dot{E}}{E}%
\right] ,  \notag \\
\dot{\sigma} &=&-3H\sigma +\frac{1}{3}E^{2}+\frac{\lambda }{6}\left(
H+\sigma \right) \left( H+7\sigma \right) E^{2}  \label{eq:shear} \\
&&-\frac{\lambda E^{2}}{6}\left[ \lambda \left( H+\sigma \right) ^{2}E^{2}+%
\frac{1}{2}E^{2}-\left( \gamma -1\right) \rho +4\left( H+\sigma \right) 
\frac{\dot{E}}{E}\right] .  \notag
\end{eqnarray}%
We have already reshuffled the Einstein equations to put them into a convenient
form. Eqs.(\ref{eq:Friedmann}) and (\ref{eq:raychaudhuri}) correspond to the
Friedmann and Raychaudhuri equations, respectively. Eq.(\ref{eq:vector}) can
be written as 
\begin{equation}
\dot{E}=-\frac{2\left( H+\sigma \right) }{1+2\lambda \left( H+\sigma \right)
^{2}}\left[ 1+2\lambda \left( H+\sigma \right) ^{2}+2\lambda \left( \dot{H}+%
\dot{\sigma}\right) \right] E.  \label{eq:electric}
\end{equation}%
The fluid obeys the usual adiabatic decay law: 
\begin{equation}
\dot{\rho}=-3\gamma \dot{\alpha}\rho .  \label{eq:fluid}
\end{equation}%
Equations (\ref{eq:Friedmann}) - (\ref{eq:fluid}) form a closed set of
non-linear ordinary differential equations. The Friedmann equation (\ref%
{eq:Friedmann}) measures the dynamical significance of each matter
component.

To gain an insight into the effect of the Horndeski's extra non-minimal
coupling, let us assume that Horndeski's modification term is dominant, that
is 
\begin{equation*}
\left\vert \lambda \left( H+\sigma \right) ^{2}E^{2}\right\vert \gg
E^{2},\rho . 
\end{equation*}%
The Friedmann equation (\ref{eq:Friedmann}) becomes 
\begin{equation}
H^{2}-\sigma ^{2}\sim \lambda \left( H+\sigma \right) ^{2}E^{2},
\label{eq:approx1}
\end{equation}%
which means that the universe must be strongly anisotropic when $\lambda <0$%
. Using the same approximation, (\ref{eq:electric}) reduces to 
\begin{equation*}
\frac{\dot{E}}{E}\sim -\frac{1-2\lambda \sigma \left( H+\sigma \right) }{%
\lambda \left( H+\sigma \right) }. 
\end{equation*}%
Now, (\ref{eq:raychaudhuri}) becomes 
\begin{equation*}
\dot{H}+H^{2}\sim -2\sigma ^{2}, 
\end{equation*}%
which is equivalent to the usual Einstein equation in an empty Bianchi I
universe dominated by the shear because the back-reaction of the Horndeski
term exactly cancels out at leading order. The same is true in the shear
evolution equation, as (\ref{eq:shear}) yields 
\begin{equation*}
\dot{\sigma}\sim -3H\sigma , 
\end{equation*}%
so that the universe isotropises in the same way as a (flat) universe
containing only perfect fluids. We conclude that the Horndeski modification $%
\mathcal{L}_{H}$ is fairly innocuous despite the formidable appearance of
its energy-momentum tensor. It should only be able to affect the evolution
of the universe when its contribution is comparable to the conventional
Maxwell term or the matter.

\subsubsection{Occurrence of a finite-time singularity}

\label{sec:sing}

The evolution equation for electric field (\ref{eq:electric}) is rather
similar to the linearized equation yielding the solution (\ref{eq:esol1}),
although it is fully non-linear in the present setup. In fact, we can
integrate (\ref{eq:electric}) analytically and obtain 
\begin{equation}
E=\frac{E_{0}}{e^{2\left( \alpha +\beta \right) }}\frac{1}{1+2\lambda \left(
H+\sigma \right) ^{2}}.  \label{eq:esol2}
\end{equation}%
This is essentially the same as the solution in FLRW background (\ref%
{eq:esol1}). Unless there is a mechanism within (\ref{eq:Friedmann}) - (\ref%
{eq:fluid}) that prevents $(H+\sigma )^{2}$ from reaching $-1/2\lambda $,
there should be a range of initial conditions for which the system
eventually hits a singularity. Since the system comes close to this singularity
precisely when Horndeski's modified term becomes comparable to the Maxwell term 
\begin{equation*}
\left\vert \lambda \left( H+\sigma \right) ^{2}E^{2}\right\vert \sim E^{2}, 
\end{equation*}%
we may see unusual dynamical behaviours in this regime.

\begin{figure}[tbph]
\begin{center}
\includegraphics{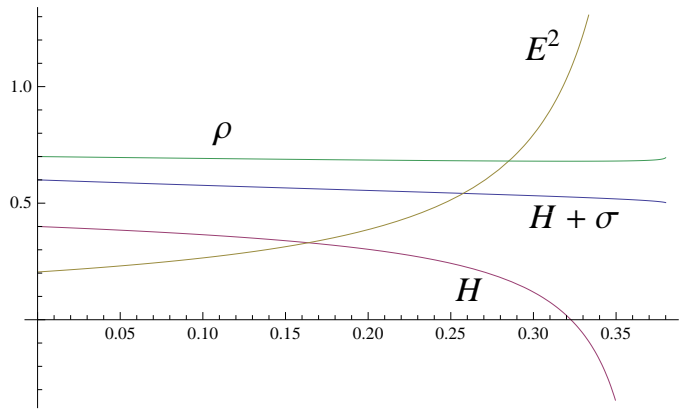}
\end{center}
\caption{The approach towards the singularity for $\protect\lambda =-2,%
\protect\gamma =0.1$ and $H+\protect\sigma >0$. The initial conditions are $%
\protect\rho =0.7,H=0.4,\protect\sigma =0.2$. The universe is initially
dominated by the matter with $p/\protect\rho =-0.9$. There is no sign of
avoiding the singularity located around $t\sim 0.38$. $H$ becomes negative
just before the singularity, which means the universe recollapses.}
\label{fig:recollapse}
\end{figure}

Let us see what happens to the evolution of the spatial geometry when the
system approaches the singularity. For this purpose, it is useful to write
down the evolution equation for $H+\sigma $, which we can cast into the
following form; 
\begin{equation}
\dot{H}+\dot{\sigma}=-\frac{1}{2}\left[ 3\left( H+\sigma \right) ^{2}+\left(
\gamma -1\right) \rho \right] +\frac{1+2\lambda \left( H+\sigma \right) ^{2}%
}{4}E^{2}.  \label{eq:key}
\end{equation}%
It is immediately clear that the right-hand side is negative definite when $%
\gamma \geq 1$ and $\lambda <0$ for initial conditions satisfying 
\begin{equation}
\left( H+\sigma \right) ^{2}>-\frac{1}{2\lambda }.  \label{eq:pathological}
\end{equation}%
Therefore, the singularity is inevitable regardless of the initial
conditions for $\gamma \geq 1$ as long as $H+\sigma >0$ is sufficiently
large initially. Although it does not apply to inflationary universes with $%
0<\gamma \ll 1$, we already know $H+\sigma $ decreases monotonically while $%
\rho $ dominates the evolution of the universe. Thus, if the condition (\ref%
{eq:pathological}) holds initially, $E$ eventually grows and any matter
domination, and hence inflation, comes to an end. Once the electric field
starts to dominate the dynamics, the right-hand side of (\ref{eq:key}) is
again negative definite and the singularity must be reached. While we are
unable to eliminate the possibility that $\dot{H}+\dot{\sigma}$ turns to
positive and the universe manages to avoid the singularity during a brief
period of matter-electric equality, numerical calculations suggest otherwise
(see figure \ref{fig:recollapse}).

%{\bf  I am not able to reproduce the remaining of this paragraph, and I am therefore not convinced that $\dot H>0$.  Instead of the equation below I get the following:
%
%\begin{equation}
%\dot{H}\sim  -\frac{2\lambda E^{2}}{3\left[ 1+2\lambda \left( H+\sigma
%\right) ^{2}\right] }\left[ 3\left( H+\sigma \right) ^{2}+\left( \gamma
%-1\right) \rho \right]  -\left( H^2+2\sigma^2+\lambda\sigma(H+\sigma)E^2 \right). 
%\end{equation}%
%
%Note that close to the singularity we have $|H|,|
%\sigma| \gg |H+\sigma|$, see figure 1.  It is therefore not clear that we can neglect the extra terms. Note that unlike $\dot H + \dot \sigma$, the Raychaudhuri and shear equation is sourced by a term proportional to the singularity ($\dot E / E$).  This indicates that the neglected terms are of the same amplitude as the others and must be taken into account.  }

\begin{figure}[tbph]
\begin{center}
\includegraphics{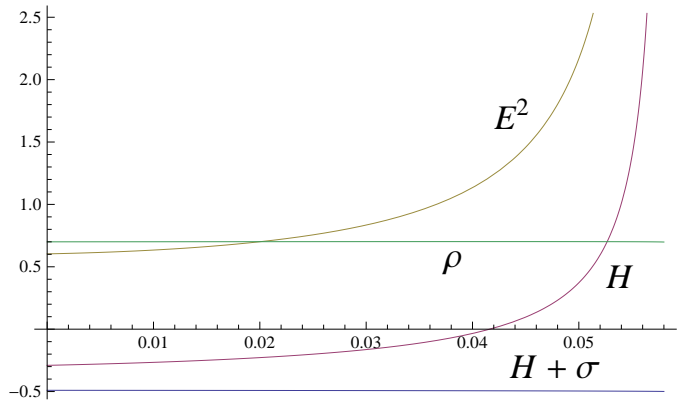}
\end{center}
\caption{The approach towards the singularity for $\protect\lambda =-2,%
\protect\gamma =0.1$ and $H+\protect\sigma <0$. The initial conditions are $%
\protect\rho =0.7,H=-0.29,\protect\sigma =-0.2$. The universe is initially
contracting, but eventually bounces and rapidly expands before it reaches
the singularity around $t\sim 0.058$.}
\label{fig:bounce}
\end{figure}

Since the instability condition (\ref{eq:pathological}) roughly corresponds
to the one for Horndeski's term to have a significant effect compared to the
Maxwell term in the Lagrangian, it effectively rules out any sensible
cosmological application of the theory with a negative $\lambda $. We have
not discussed the case of negative $H+\sigma $ since it describes either a
collapsing universe or excessively anisotropic one. Close examination of (%
\ref{eq:raychaudhuri}) indicates recollapse or bounce right before hitting
the singularity, both deriving from the violation of the weak energy
condition. These behaviours are also observed in our numerical solutions of
the equations (see figure \ref{fig:bounce}). 

\subsection{Expansion-normalised autonomous system}

In the previous subsection, we saw that the theory with a negative coupling, 
$\lambda $, should lead to pathological dynamics when the Horndeski
modification has an appreciable effect. On the other hand, when $\lambda $
is positive there is no danger of a singularity. The energy density is
positive definite and we expect some viable cosmological dynamics. In order
to carry out a more systematic investigation, it is always useful to rewrite
the equations in terms of density parameters defined for each matter
component, including the Horndeski contribution. It also enables us to apply
the conventional methods of dynamical systems analysis.

We introduce the following normalised variables: 
\begin{equation*}
\Sigma =\frac{\sigma }{H},\ \ \ \ \ \Omega _{M}=\frac{E^{2}}{6H^{2}},\ \ \ \
\ \Omega _{H}=6\lambda \left( H+\sigma \right) ^{2}\Omega _{M},\ \ \ \ \
\Omega _{m}=\frac{\rho }{3H^{2}}, 
\end{equation*}%
where $\Omega _{H}\gtrless 0$ correspond to $\lambda \gtrless 0$
respectively. The normalised Friedmann equation 
\begin{equation}
1=\Sigma ^{2}+\Omega _{H}+\Omega _{M}+\Omega _{m},  \label{eq:Friedmann2}
\end{equation}%
will be used as the standard measure of the dynamical significance of each
component. In particular, when $\Omega _{H}>0$, all the parameters are
bounded by $1$ so that their values have a clear physical interpretation. Note that $\Omega_i=\rho_i/3H^2$ for $i=(M,H)$, where $\rho_M$ and $\rho_H$ represent the minimal and non-minimal contribution to the vector field energy density, respectively. 
From (\ref{eq:raychaudhuri}), we define the deceleration parameter as 
\begin{eqnarray}
q &=&-\frac{\dot{H}}{H^{2}}-1  \notag \\
&=&2\Sigma ^{2}+\Omega _{M}+\frac{3\gamma -2}{2}\Omega _{m}-\frac{\left(
\Omega _{M}-\Omega _{H}\right) \Omega _{H}}{2\left( 1+\Sigma \right) ^{2}}%
\left( 1-\frac{3(\gamma -1)\Omega _{m}}{3\Omega _{M}+\Omega _{H}}\right)
\label{eq:deceleration} \\
&&+\frac{\Omega _{H}}{2\left( 1+\Sigma \right) \left( 3\Omega _{M}+\Omega
_{H}\right) }\left( 3\left( 3+5\Sigma \right) \Omega _{M} - \left( 1-\Sigma
\right) \Omega _{H}\right) .  \notag
\end{eqnarray}%
Following the standard method \cite{Wainwright}, we switch the time
coordinate from $t$ to $\alpha $. Using (\ref{eq:Friedmann2}) and (\ref%
{eq:deceleration}), we derive the following evolution equations for the
expansion-normalised variables: 
\begin{eqnarray}
\frac{d\Sigma }{d\alpha } &=&\left( q-2\right) \Sigma -2\Sigma ^{2}+\Omega
_{M}-\frac{3\gamma -2}{2}\Omega _{m}+q,  \label{eq:Sigma1} \\
\frac{d\Omega _{M}}{d\alpha } &=&2\Omega _{M}\left[ q+1-\frac{\Omega _{H}}{%
1+\Sigma }\left( 1-\frac{3(\gamma -1)\Omega _{m}}{3\Omega _{M}+\Omega _{H}}%
\right) -\frac{6\Omega _{M}-\Omega _{H}}{3\Omega _{M}+\Omega _{H}}\left(
1+\Sigma \right) \right] ,  \label{eq:Maxwell1} \\
\frac{d\Omega _{H}}{d\alpha } &=&2\Omega _{H}\bigg[q-2\Sigma +\frac{\Omega
_{M}-\Omega _{H}}{1+\Sigma }-\frac{6\Omega _{M}-\Omega _{H}}{3\Omega
_{M}+\Omega _{H}}\left( 1+\Sigma \right)  \label{eq:Horndeski1} \\
&&\ -\frac{\Omega _{m}}{1+\Sigma }\left( \frac{3\gamma -2}{2}-\frac{3\left(
\gamma -1\right) \Omega _{H}}{3\Omega _{M}+\Omega _{H}}\right) \bigg], 
\notag \\
\frac{d\Omega _{m}}{d\alpha } &=&\left( 2q-3\gamma +2\right) \Omega _{m}.
\label{eq:matter}
\end{eqnarray}%
These four equations are not independent since they are related by the first
integral (\ref{eq:Friedmann2}).

\subsubsection{Fixed points in the dynamical system}

We first classify the fixed points. Since the subsystem specified by $\Omega
_H =0$ is identical to the magnetic Bianchi type I discussed in \cite%
{collins72, blanc97}, we know there must be at least four fixed points of
physical interest:

\begin{description}
\item[ Flat Friedmann universe : $F$ ] 
\begin{equation*}
\left( \Sigma , \Omega _M , \Omega _H , \Omega _m \right) = \left( 0 , 0, 0,
1 \right) . 
\end{equation*}

\item[ Electric Bianchi type I : $E$ ] 
\begin{equation*}
\left( \Sigma , \Omega _M , \Omega _H , \Omega _m \right) = \left( \frac{%
3\gamma -4}{4}, \frac{3}{16} \left( 2-\gamma \right) \left( 3\gamma -4
\right) ,0 , \frac{3}{8}(4 -\gamma ) \right) . 
\end{equation*}
The existence condition is $\gamma > 4/3$.

\item[ Kasner solutions : $K_{\pm }$ ] 
\begin{equation*}
\left( \Sigma , \Omega _M , \Omega _H , \Omega _m \right) = \left( \pm 1 ,
0, 0 ,0 \right) . 
\end{equation*}
\end{description}

In addition, there appears a fixed point describing a universe dominated by
the Horndeski energy density:

\begin{description}
\item[ $\Omega _H$-dominated universe : $H_E$ ] 
\begin{equation*}
\left( \Sigma , \Omega _M , \Omega _H , \Omega _m \right) = \left( 0,0,1,0
\right) . 
\end{equation*}
The deceleration parameter for this solution is $q = 0$, which is consistent with the analysis in section \ref{chsub:electric} where we showed that the back-reaction of the vector field exactly cancels out to leading order when the energy density is dominated by Horndeski's modification term.
\end{description}

There are subtleties regarding these fixed points. First of all, $E$ and $%
H_{E}$ do not represent physical space-times when $\lambda \neq 0$ since
they imply $H=\sigma =0$ and $(H+\sigma )^{-1}=0$ respectively. It does not
mean these fixed points are irrelevant in the dynamics, however, since they
may be reached asymptotically from finite $H$ and $\sigma $ in the far past
or future. We shall see an example of this in figure \ref{fig:3Delectric}.
Secondly, $F$ and $K_{\pm }$ must be treated with care since some of the
denominators appearing in the evolution equations (\ref{eq:Sigma1})-(\ref%
{eq:matter}) vanish on those fixed points. Nevertheless, it does not mean they are
unphysical since they are well-behaved when appropriate limits are taken for
the numerators. But the analysis requires evaluation of $0/0$, which implies
the stabilities may depend on the way the fixed point is approached. It is a
consequence of the fact that $H$ is not decoupled from the normalised
variables, and implicitly appears in the definition of $\Omega _{H}$. While
a fixed point in the expansion-normalised equations usually represents a
self-similar solution that is invariant under a scale transformation, the
dynamical effect of the Horndeski modification depends on the scale of the
curvature, or the size of the universe. Therefore, it is not surprising to
see the stability change depending on each orbit with its specific value of $%
H$. We will find this is indeed the case.

\subsection{Dynamics around the matter-dominated solution}

\label{sec:electricF} From physical point of view, by far the most
interesting fixed point is $F$ because it can be regarded as a model of the
late-time evolution for the universe when $\gamma =1$ (dust) or $\gamma =4/3$
(radiation), and also a model of inflation when $\gamma <2/3$. We have already
mentioned the instability against perturbations of the electric field for $%
\lambda <0$. The condition for the occurrence of a singularity (\ref{eq:pathological}) translates into 
\begin{equation}
3\Omega _{M}+\Omega _{H}<0 \label{eq:pathological2}
\end{equation}%
in the new variables. Here, we shall see that this condition coincides with
the instability condition for $F$ and otherwise the dynamics is trivial. We
also study the stability for positive $\lambda $ and show that it depends on
the value of $H$.

\subsubsection{Linearisation}

For the purpose of linearisation around $F$, it turns out to be convenient
to eliminate $\Omega _{m}$ using the Friedmann equation (\ref{eq:Friedmann2}%
) and rewrite the equations as 
\begin{eqnarray}
\frac{d\Sigma }{d\alpha } &=&f_{\Sigma }(\Sigma ,\Omega _{M},\Omega _{H}), \\
\frac{d\Omega _{M}}{d\alpha } &=&f_{M}(\Sigma ,\Omega _{M},\Omega _{H}), \\
\frac{d\Omega _{H}}{d\alpha } &=&f_{H}(\Sigma ,\Omega _{M},\Omega _{H}),
\end{eqnarray}%
whose right-hand sides we avoid writing down explicitly as they are lengthy.
While there are apparent $0/0$s in those equations when they are evaluated
on $F$, they should be all well defined if an appropriate limit is taken
along an arbitrary reference orbit. To proceed, we evaluate the functions $%
f_{i},i=\Sigma ,M,H$ for $\Sigma =0$ and then take the limit $(\Omega
_{M},\Omega _{H})\rightarrow (0,0)$: 
\begin{eqnarray*}
\lim_{(\Omega _{M},\Omega _{H})\rightarrow (0,0)}f_{\Sigma }(0,\Omega
_{M},\Omega _{H}) &=&-2\gamma L_{2}, \\
\lim_{(\Omega _{M},\Omega _{H})\rightarrow (0,0)}f_{M}(0,\Omega _{M},\Omega
_{H}) &=&-2\gamma L_{2}, \\
\lim_{(\Omega _{M},\Omega _{H})\rightarrow (0,0)}f_{H}(0,\Omega _{M},\Omega
_{H}) &=&6\gamma L_{2},
\end{eqnarray*}%
where we have introduced a notation 
\begin{equation*}
L_{n}=\lim_{(\Omega _{M},\Omega _{H})\rightarrow (0,0)}\frac{(\Omega _{H})^{n}%
}{3\Omega _{M}+\Omega _{H}}, 
\end{equation*}%
which will also be used later. When $\Omega _{H}>0$, or equivalently $%
\lambda >0$, we have $L_{2}=0$ and the fixed point $F$ is always well
defined as it should be. The existence of the limit is inconclusive when $%
\Omega _{H}<0$ ($\lambda <0$) and the orbit satisfies 
\begin{equation}
\lim_{(\Omega _{M},\Omega _{H})\rightarrow (0,0)}\frac{\Omega _{H}}{\Omega
_{M}}=-3.  \label{eq:singlimit}
\end{equation}%
However, such an orbit merely represents one that ends up in the singularity 
$2\lambda (H+\sigma )^{2}=-1$. Since we have already discussed this case in
detail, we exclude those orbits from our consideration here. As long as an
orbit does not hit the singularity when approaching $F$, $L_{2}$ should
exist and be equal to zero.

Since the right-hand sides can be evaluated only as a limit associated with
each reference orbit, the linearisation takes an extra step. We first expand
the equations around an arbitrary point $(\Sigma _{0},\Omega _{M0},\Omega
_{H0})$ and then take the limit $(\Sigma _{0},\Omega _{M0},\Omega
_{H0})\rightarrow (0,0,0)$. The resultant linear equations are given as
follows: 
\begin{eqnarray}
\frac{d\delta \Sigma }{d\alpha } &=&\frac{3}{2}(\gamma -2)\delta \Sigma
+\left( 2+6\gamma L_{1}^{2}\right) \delta \Omega _{M}+\frac{1}{2}\left[
\gamma +2+4\gamma L_{1}\left( L_{1}-2\right) \right] \delta \Omega _{H},
\label{eq:lS1} \\
\frac{d\delta \Omega _{M}}{d\alpha } &=&\left( 3\gamma -4+6\gamma
L_{1}^{2}\right) \delta \Omega _{M}+2\gamma \left( L_{1}-1\right) ^{2}\delta
\Omega _{H},  \label{eq:lM1} \\
\frac{d\delta \Omega _{H}}{d\alpha } &=&-18\gamma L_{1}^{2}\delta \Omega
_{M}-2\left[ 2+3\gamma L_{1}\left( L_{1}-2\right) \right] \delta \Omega _{H},
\label{eq:lH1}
\end{eqnarray}%
where the $\delta $s preceding the variables denote their small
perturbation. Unless the orbit is the singular one specified by (\ref%
{eq:singlimit}), $L_{1}$ is finite and therefore these linearised equations
are well defined. However, the asymptotic value of $L_{1}$ does depend on
each orbit. Going back to its definition, one notices that 
\begin{equation*}
L_{1}=\frac{\mathcal{R}}{\mathcal{R}+3}, 
\end{equation*}%
where 
\begin{equation*}
\mathcal{R}=\frac{\Omega _{H}}{\Omega _{M}}=6\lambda (H+\sigma )^{2}. 
\end{equation*}%
We already know its behaviour near $F$ since we have 
\begin{equation*}
\mathcal{R}\sim 6\lambda H^{2}\sim 2\lambda \rho . 
\end{equation*}%
Solving (\ref{eq:fluid}), we obtain 
\begin{equation}
\mathcal{R}=\mathcal{R}_{0}e^{-3\gamma \alpha },  \label{eq:r}
\end{equation}%
where $\mathcal{R}_{0}$ is an orbit-specific constant. We also note that 
\begin{equation}
\delta \Omega _{H}=\mathcal{R}\delta \Omega _{M}  \label{eq:lconstraint}
\end{equation}%
along each orbit, so that we have an additional linear constraint. Now, the
linearized equations (\ref{eq:lS1})-(\ref{eq:lH1}) can be written 
\begin{align}
\frac{d\delta \Sigma }{d\alpha }& =\frac{3}{2}(\gamma -2)\delta \Sigma +%
\frac{1}{3+\mathcal{R}}\left( 6+\frac{1}{2}\mathcal{R}\big[10+3\gamma -%
\mathcal{R}(3\gamma -2)\big]\right) \delta \Omega _{M},  \label{eq:lS2} \\
\frac{d\delta \Omega _{M}}{d\alpha }& =\delta \Omega _{M}\left[ 3\gamma -4+%
\frac{6\gamma \mathcal{R}}{3+\mathcal{R}}\right] ,  \label{eq:lM2} \\
\frac{d\delta \Omega _{H}}{d\alpha }& =\delta \Omega _{H}\left[ -4+\frac{%
6\gamma \mathcal{R}}{3+\mathcal{R}}\right] .  \label{eq:lH2}
\end{align}%
and we can easily read off the eigenvalues of the linearization matrix 
\begin{equation}
\left( \frac{3}{2}(\gamma -2),3\gamma -4+\frac{6\gamma \mathcal{R}}{\mathcal{%
R}+3},-4+\frac{6\gamma \mathcal{R}}{\mathcal{R}+3}\right) .
\label{eq:eigenvalues}
\end{equation}%
Notice that the eigenvalues are orbit and time dependent through $\mathcal{R}%
(\alpha )$. As it was necessary to take a non-standard approach to obtain
the eigenvalues, we will later confirm the validity of the result by
performing numerical calculations.

\subsubsection{Stability of the matter-dominated solution}

The first of the eigenvalues (\ref{eq:eigenvalues}) is negative and
represents the stability of $F$ against perturbation of $\Sigma $. Since the
third eigenvalue is always smaller than the second, the condition for the
stability is 
\begin{equation*}
3\gamma -4 + \frac{6\gamma \mathcal{R}}{\mathcal{R}+3} < 0. 
\end{equation*}

Let us first consider $\lambda >0,$ which corresponds to $\mathcal{R}>0$. In
this case, we have 
\begin{equation*}
0<\frac{\mathcal{R}}{\mathcal{R}+3}<1, 
\end{equation*}%
and consequently $F$ is definitely stable for $\gamma <4/9$, which means
positive $\lambda $ cannot be relevant in the context of inflation, and $F$
is unstable for $\gamma >4/3$. For $4/9<\gamma <4/3$, the stability is
orbit-dependent. When an orbit satisfies 
\begin{equation}
\mathcal{R}>\frac{12-9\gamma }{9\gamma -4},  \label{eq:instability}
\end{equation}%
it runs away from $F$. When $\mathcal{R}$ is smaller than this threshold
value, the orbit is attracted towards $F$. Note that the value of $\mathcal{R%
}$ is time-dependent so that the stability can change over the course of the
evolution. In particular, from (\ref{eq:r}), $\mathcal{R}$ is
monotonically decreasing as long as the orbit stays close to the
matter-dominated solution. We can immediately conclude that for orbits with $%
\mathcal{R}<(12-9\gamma )/(9\gamma -4)$, the stability does not change as
the universe expands. For those satisfying (\ref{eq:instability}), they
typically become stable asymptotically in the future since $\mathcal{R}$ can
only increase when the universe is dominated by both Maxwell's and
Horndeski's terms. For a physically interesting range of initial conditions,
we shall later confirm that the instability of the electric field saturates
before the orbit goes too far from $F$ and eventually comes back to it.

\begin{figure}[tbph]
\begin{center}
\includegraphics[width=0.5\textwidth]{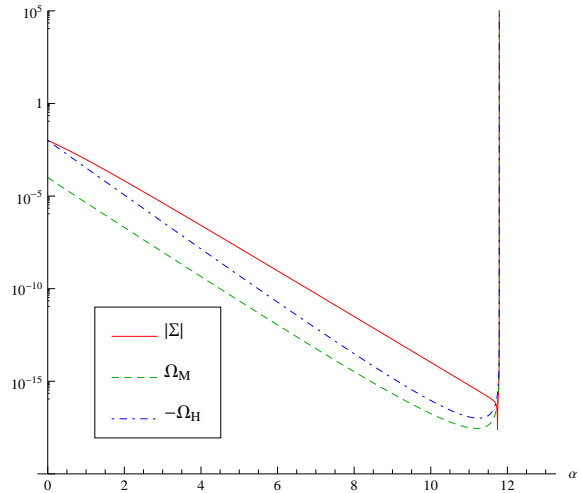}
\end{center}
\caption{Simulation of (\protect\ref{eq:Sigma1})-(\protect\ref{eq:Horndeski1}%
) for an inflationary universe with $\protect\gamma =0.1$ and initial
conditions close to $F$ ($\Sigma =-0.01$, $\Omega _{M}=0.0001$, $\Omega
_{H}=-0.01$). The singularity at $3\Omega _{M}+\Omega _{H}=0$ occurs at time 
$\protect\alpha =11.65$.}
\label{fig:singularityF}
\end{figure}

For $\lambda <0$, the dynamics is very different -- depending on $\mathcal{R}%
\gtrless -3$. As was already mentioned, the critical value $\mathcal{R}=-3$
corresponds to the singularity $2\lambda (H+\sigma )^{2}=-1$ discussed
in section \ref{sec:sing}. Firstly, $\mathcal{R}\in (-3,0)$ implies $\mathcal{R}/(\mathcal{R%
}+3)\in (-\infty ,0)$ and therefore the orbits in this range are stable as
long as $\gamma <4/3$. For $\mathcal{R}<-3$, we have $\mathcal{R}/(\mathcal{R%
}+3)\in (1,\infty )$ and $\mathcal{R}$ monotonically increases in the
vicinity of $F$. It approaches $\mathcal{R}=-3$ from below and therefore any
orbit eventually becomes unstable. Since we already know $\mathcal{R}=-3$ is
the singularity, we conclude that the orbits with this range of initial $%
\mathcal{R}$ can never settle down at $F$, regardless of the equation of
state parameter $\gamma $; see figure \ref{fig:singularityF} for simulation
of an inflationary universe with initial conditions satisfying $\mathcal{R}%
=-100$. The solution approaches $F$ until $\mathcal{R}$ is close to the
critical value in which case the universe moves away from $F$. 
After leaving $F$, we expect the system enters the regime where $\Omega _{m}$
is dynamically negligible. Then the analysis in the section \ref{sec:sing}
applies and the singularity is inevitable.

\subsubsection{Dynamics with $\protect\lambda >0$ in a radiation- or
dust-dominated universe \label{ch:lineardynamics}}

We have found that the theory is quite innocuous for positive non-minimal
coupling constant, $\lambda ,$ in which case there is no singularity and the
Friedmann solution $F$ is stable at late times. However, as shown above,
there is a transient period where $F$ is unstable for fluids satisfying $%
\gamma >4/9$. In this period, $\Omega _{M}$ and $\Omega _{H}$ grow and, if
the instability does not saturate before they become too large, the universe
will eventually become strongly anisotropic. Since this introduces potential
problems in the radiation or dust-dominated epoch, it is of interest to
specify the range of initial conditions such that the universe is close to $F
$ at all the subsequent times, i.e., $|\Omega _{m}-1|\ll 1$. In this
subsection, therefore, we  investigate the dynamics close to $F$ in more
detail for dust ($\gamma =1$) and radiation ($\gamma =4/3$) with a positive
non-minimal coupling constant ($\lambda >0\Leftrightarrow \Omega _{H}>0$).

\begin{figure}[htbp]
\begin{center}
\includegraphics[width=1.0\textwidth]{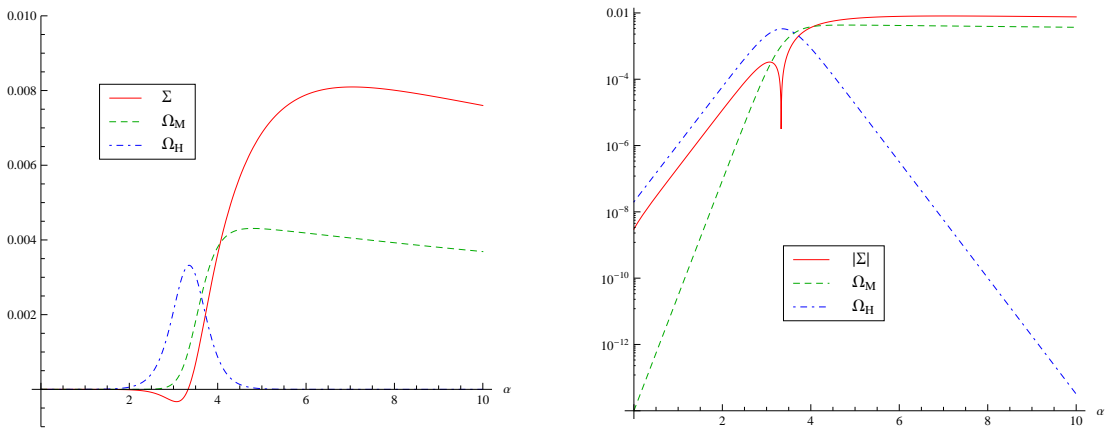}
\end{center}
\caption{Simulation of (\protect\ref{eq:Sigma1})-(\protect\ref{eq:Horndeski1}%
) for radiation ($\protect\gamma=4/3$) and a positive non-minimal coupling
coupling constant ($\protect\lambda>0$). Initial conditions ($\Sigma=-3 *
10^{-9}$, $\Omega_M=10^{-14}$, $\Omega_H=2*10^{-8}$) are such that the orbit
is always close to the Friedmann solution $F$. From the logarithmic plot to
the right it is clear that the ratio $\Omega_H/\Omega_M$ is monotonically
decaying in agreement with equation (\protect\ref{eq:r}).}
\label{fig:CloseToRadiation}
\end{figure}

We can easily solve the linearized equations (\ref{eq:lM2}) and (\ref{eq:lH2}%
) exactly: 
\begin{align}
\delta \Omega _{M}(\alpha )& =\delta \Omega _{M0}\left( \frac{3+\mathcal{R}%
_{0}}{3+\mathcal{R}}\right) ^{2}e^{-(4-3\gamma )\alpha }, \\
\delta \Omega _{H}(\alpha )& =\delta \Omega _{H0}\left( \frac{3+\mathcal{R}%
_{0}}{3+\mathcal{R}}\right) ^{2}e^{-4\alpha },
\end{align}%
where $\delta \Omega _{M0}$ and $\delta \Omega _{H0}$ are integration
constants. At late times, when $\mathcal{R}\propto e^{-3\gamma \alpha
}\rightarrow 0$, the Maxwell and Horndeski densities decay as $\delta \Omega
_{M}\propto e^{-(4-3\gamma )\alpha }$ and $\delta \Omega _{H}\propto
e^{-4\alpha }$. This is consistent with $F$ being an attractor at late times
as shown above. When $\mathcal{R}\gg 3,$ the Maxwell and Horndeski densities
grow as $\delta \Omega _{M}\propto e^{(9\gamma -4)\alpha }$ and $\delta
\Omega _{H}\propto e^{2(3\gamma -2)\alpha }$. We note that the maximum
values of $\delta \Omega _{M}$ and $\delta \Omega _{H}$ occur when $\mathcal{%
R}=(12-9\gamma )/(-4+9\gamma )$ and $\mathcal{R}=6/(3\gamma -2)$,
respectively. Taking into account the linear constraint (\ref{eq:lconstraint}%
), it follows that the maximum values of both $\Omega _{M}$ and $\Omega _{H}$
are roughly equal. Now we find that the initial conditions must satisfy 
\begin{equation}
\delta \Omega _{H0}\ll \left( \delta \Omega _{M0}\right) ^{\frac{1}{2}}\quad
\Leftrightarrow \quad \lambda \ll \left( E_{0}H_{0}\right) ^{-1}
\label{eq:conditionRadiation}
\end{equation}%
in a radiation-dominated universe and 
\begin{equation}
\delta \Omega _{H0}\ll \left( \delta \Omega _{M0}\right) ^{\frac{2}{5}}\quad
\Leftrightarrow \quad \lambda \ll \left( E_{0}\right) ^{-\frac{6}{5}}\left(
H_{0}\right) ^{-\frac{4}{5}}  \label{eq:conditionDust}
\end{equation}%
in a dust dominated universe to ensure that $|\Omega _{m}-1|\ll 1$ at all
the subsequent times. In figures \ref{fig:CloseToRadiation} and \ref%
{fig:CloseToDust}, we show a numerical integration of the full non-linear
equations (\ref{eq:Sigma1})-(\ref{eq:Horndeski1}) for radiation and dust,
respectively. Since the initial conditions just barely satisfy the
conditions (\ref{eq:conditionRadiation})-(\ref{eq:conditionDust}), the peak
values of $\Omega _{H}$ and $\Omega _{M}$ are at the one-percent level of the total energy budget. Note
that, to good accuracy, we have $\Sigma \lessgtr 0$ when $\Omega
_{M}\lessgtr \Omega _{H}$. This introduces the possibility of cancelling the
effects of spatial anisotropy on the Cosmic Microwave Background (CMB),
which will be discussed in section \ref{ch:constraints}. Under the
assumption that the theory describes a generalised electrodynamics, we show
in section \ref{ch:constraints} that the amplification of the electric field
must come to an end before the start of big bang nucleosynthesis, when the
temperature is $T\simeq 1 $ MeV. This still leaves the possibility open for a
huge amplification of electric fields in the period between inflationary
reheating and the nucleosynthesis. In the period of amplification, the
electric field grows very quickly, $\Omega _{M}\propto e^{8\alpha }$. After
the peak value is reached, $\Omega _{H}$ rapidly decays and soon becomes
negligible. At that stage, the dynamics becomes similar to the conventional
electrodynamics; $\Omega _{M}$ decays logarithmically (constant at the
linear level) until the dust-dominated epoch when it decays as $\Omega
_{M}\propto e^{-\alpha }$.

\begin{figure}[htbp]
\begin{center}
\includegraphics[width=1.0\textwidth]{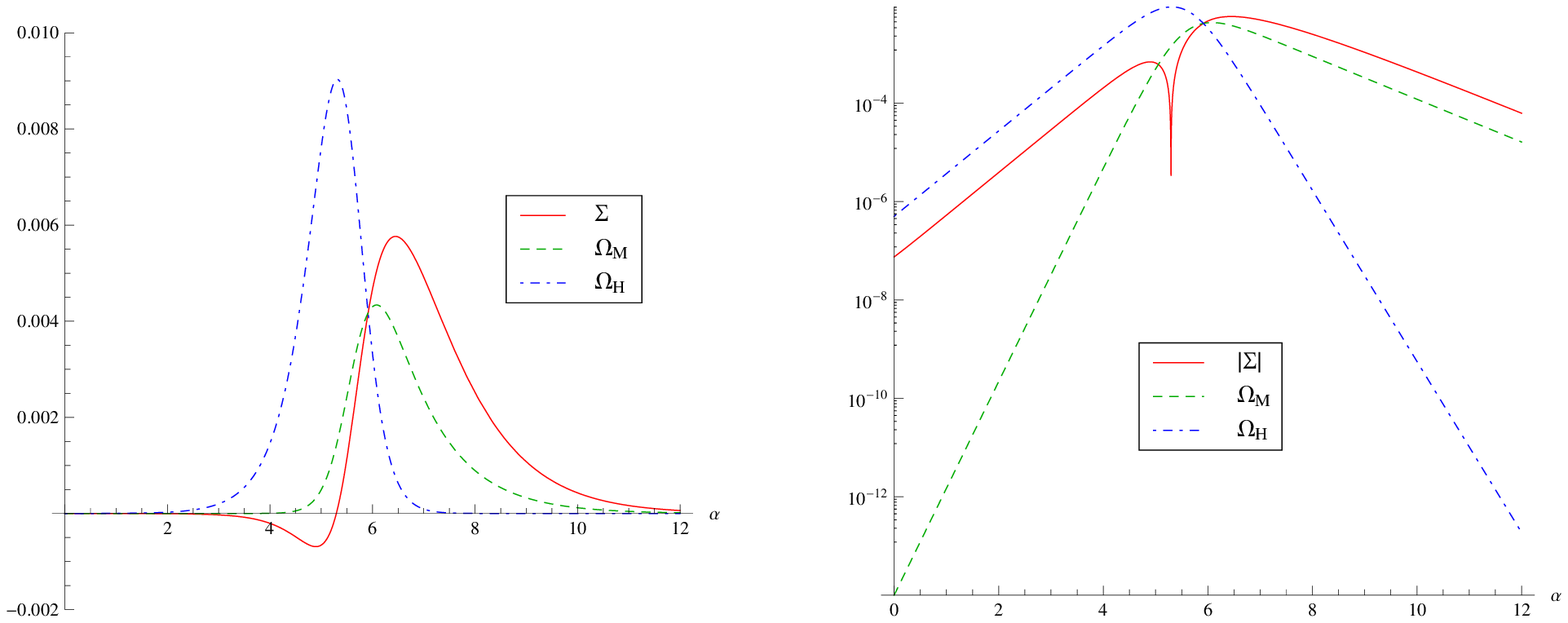}
\end{center}
\caption{Simulation of (\protect\ref{eq:Sigma1})-(\protect\ref{eq:Horndeski1}%
) for dust ($\protect\gamma=1$) and a positive non-minimal coupling coupling
constant ($\protect\lambda>0$). Initial conditions ($\Sigma=-7.5 * 10^{-8}$, 
$\Omega_M=10^{-14}$, $\Omega_H=5*10^{-7}$) are such that the orbit is always
close to the Friedmann solution $F$. From the logarithmic plot to the right
it is clear that the ratio $\Omega_H/\Omega_M$ is monotonically decaying in
agreement with equation (\protect\ref{eq:r}).}
\label{fig:CloseToDust}
\end{figure}

\subsection{Stability of the other fixed points}

Given the complexity of the dynamics, it is also helpful to analyse the
stability of the other fixed points. Here we present the eigenvalues for $%
K_{+}$, $M$ and $H_{E}$. The linear stability analysis for $K_{-}$ is
inconclusive since we have $H+\sigma =0,E=0$ there, and so the Horndeski
energy density $3\lambda (H+\sigma )^{2}E^{2}$ is generically second order in
perturbation. Our numerical simulations in the next subsection indicate that 
$K_{-}$ is a past attractor for $\lambda >0$.

When either $\Omega _M $ or $\Omega _H$ is nonzero, the stability analysis
is straightforward since there is no orbit-dependence. We obtain the
following eigenvalues:

\begin{description}
\item[ Fixed point $M$ ] 
\begin{equation*}
\left( -\frac{3}{4} \left[ 2 - \gamma \pm \sqrt{(2-\gamma ) ( 3\gamma ^2
-17\gamma +18)} \right] , -3\gamma \right) , 
\end{equation*}

\item[ Fixed point $H_E$ ] 
\begin{equation*}
\left( \pm 2 , 2 -3\gamma \right) . 
\end{equation*}
\end{description}

Whenever $M$ exists ($\gamma >4/3$), it is a future attractor. The new fixed
point $H_E$ is a saddle regardless of $\gamma $, and the orbit is
temporarily attracted towards it for initial conditions close to $F$ that
fail to satisfy the conditions (\ref{eq:conditionRadiation})-(\ref%
{eq:conditionDust}). We did not find the dynamics around this solution to be
of any phenomenological interest.

The eigenvalues of $K_{+}$ is dependent on each reference orbit and the
linearisation can be carried out in a similar way as for $F$ in section \ref{sec:sing}. In the end, we
derive the following equations:
\begin{eqnarray*}
\frac{d\delta \Sigma }{d\alpha } &=&3(2-\gamma )\delta \Sigma +\frac{%
24-9\gamma +\mathcal{R}\left[ 2\mathcal{R}+38-3\gamma (\mathcal{R}+4)\right] 
}{2(\mathcal{R}+3)}\delta \Omega _{M}, \\
\frac{d\delta \Omega _{M}}{d\alpha } &=&\frac{10\mathcal{R}-6}{\mathcal{R}+3}%
\delta \Omega _{M}, \\
\frac{d\delta \Omega _{H}}{d\alpha } &=&\frac{4\mathcal{R}-24}{\mathcal{R}+3}%
\delta \Omega _{H},
\end{eqnarray*}%
and read off the eigenvalues: 
\begin{equation*}
\left( 3(2-\gamma ),\frac{10\mathcal{R}-6}{\mathcal{R}+3},\frac{4\mathcal{R}%
-24}{\mathcal{R}+3}\right) .
\end{equation*}%
The first eigenvalue is associated with the perturbation of $\Omega _{m}$
and positive so that $K_{+}$ cannot be a future attractor. The
time-dependence of $\mathcal{R}=6\lambda (H+\sigma )^{2}$ around $K_{+}$ is
given by evaluating (\ref{eq:key}) with $H\sim \sigma $ as 
\begin{equation*}
\mathcal{R}\sim \mathcal{R}_{0}e^{-6\alpha }.
\end{equation*}%
For $\mathcal{R}<0$, the situation is analogous to $F$ except that $K_{+}$
is irrelevant to future asymptotic behaviour. An orbit with $\mathcal{R}<-3$
initially reaches the singularity while $K_{+}$ is a saddle point for
the others. For $\mathcal{R}>0$, the stability changes over the course of
the evolution.

%Finally, we conclude this section by demonstrating the past stability of $K_-$ for $\lambda >0$
%numerically. Fig.\ref{fig.phaseE} shows the phase portrait of the invariant subset
%$\Omega _m =0$ (a universe without the fluid).  It clearly shows the past attractor $K_-$ for $\lambda >0 \Leftrightarrow \Omega _H >0$.
%In the region $\Omega _H <0$, there are two distinct flows inside and outside the red
%line that denotes the location of the singularity $3 \Omega _M + \Omega _H =0$. 

\subsection{Numerical Analysis}

\begin{figure}[htbp]
\begin{center}
\includegraphics[width=0.8\textwidth]{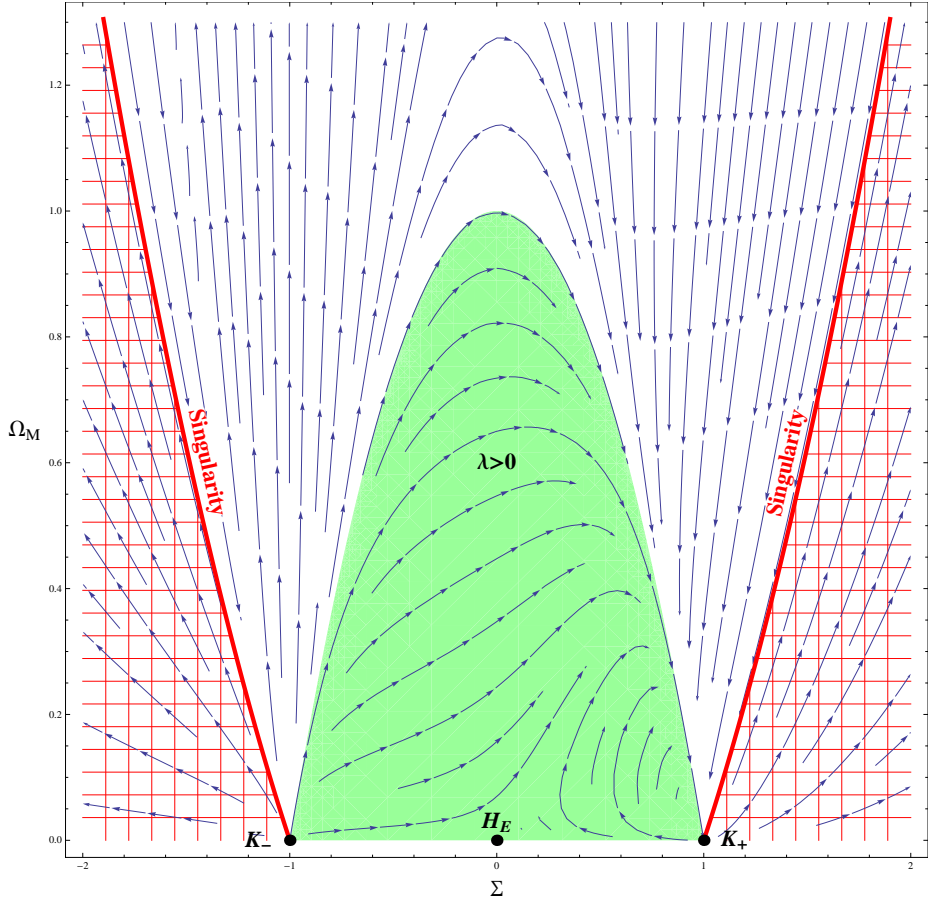}
\end{center}
\caption{Phase flow for the subsystem $\Omega_m=0$ in the electric case. The
green shaded region corresponds to $\protect\lambda>0$. The red bold line is
the position of the singularity, while the red mesh is the region where the
singularity is inevitable.}
\label{fig:phaseE}
\end{figure}

\begin{figure}[htbp]
\begin{center}
\includegraphics[width=0.5\textwidth]{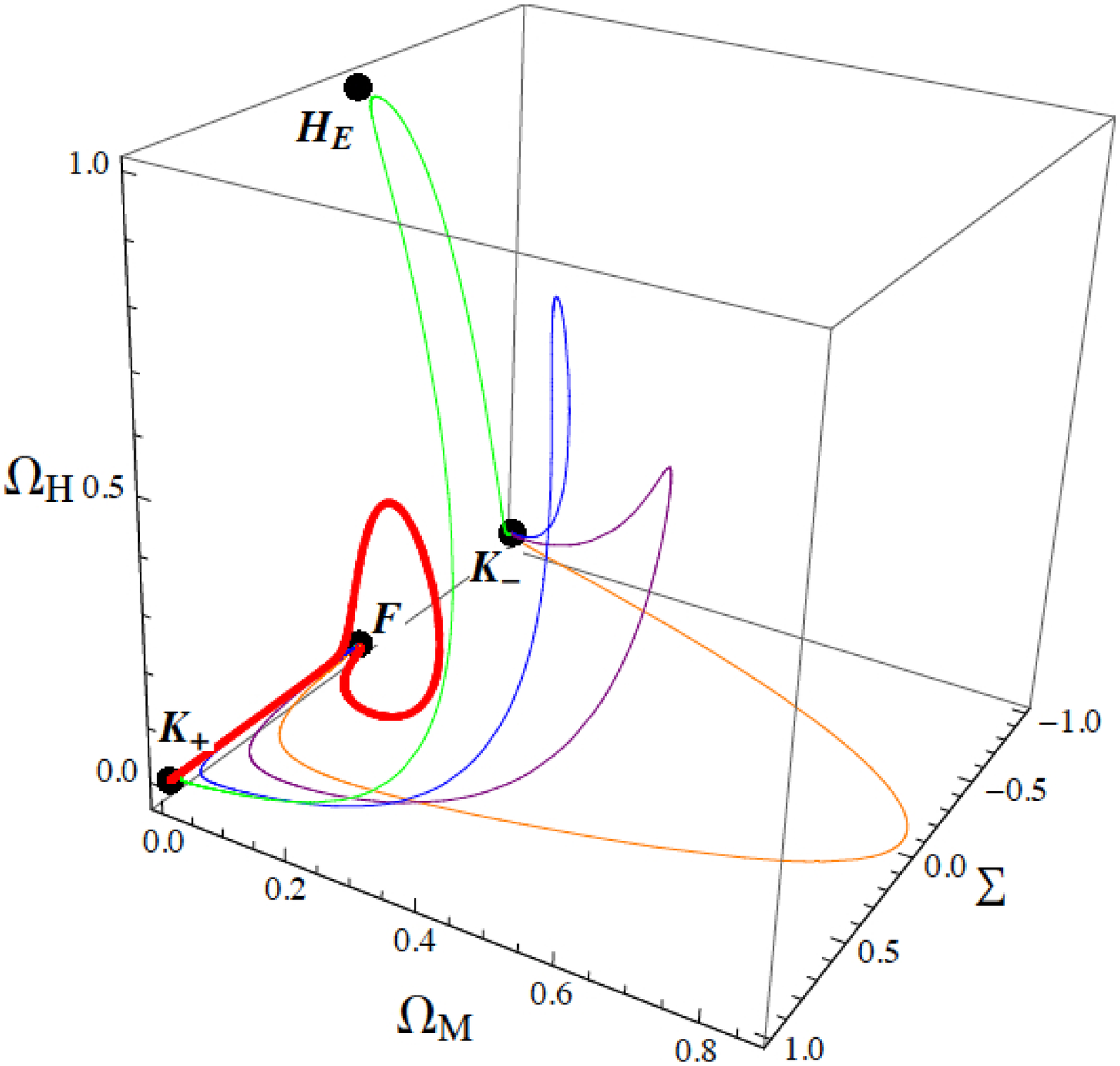}
\end{center}
\caption{ Simulations of the non-linear equations (\protect\ref{eq:Sigma1})-(%
\protect\ref{eq:Horndeski1}) for a radiation fluid ($\protect\gamma=4/3$).
All the orbits correspond to a positive non-minimal coupling, $\protect%
\lambda>0$. The thick red line starts from $K_{+}$, gets
repelled from $F$ and eventually comes back to it, which confirms the
time-dependence of the stability for $F$. All the other orbits traverse from 
$K_-$ to $F$ indicating $K_-$ is a past attractor.}
\label{fig:3Delectric}
\end{figure}

We conclude this section by showing two visualizations that confirm the
analysis of this section, and suggest that $K_{-}$ is a past attractor
(which is the case in the minimally coupled theory) for $\lambda >0$. Figure %
\ref{fig:phaseE} shows a phase portrait of the invariant subset $\Omega
_{m}=0$ (a universe without the fluid). This subset is effectively two
dimensional by the Friedmann equation (\ref{eq:Friedmann2}) and the phase
space is completely characterised by the variables $\Sigma $ and $\Omega _{M}
$. The green shaded region corresponds to $\lambda >0$. It clearly shows
that $K_{-}$ is a past attractor for $\lambda >0$ on this subspace. Note the
existence of stream lines directed towards $K_{-}$ in the $\lambda <0$
region  indicating orbit dependence of the stability for a negative
non-minimal coupling constant. In the region $\lambda <0$, there are two
distinct flows inside and outside the bold red lines that denote the
location of the singularity. The outer region is marked with a red mesh and
corresponds to the pathological region (\ref{eq:pathological2}) where the
singularity is inevitable. Note that (\ref{eq:pathological2}) can be written 
$1+2\Omega _{M}<\Sigma ^{2}+\Omega _{m}$ which explains why the entire
region $\Sigma ^{2}<1$ is non-pathological on the subspace $\Omega _{m}=0$.
When a fluid is included, however, the singularity can be reached even from
a small initial shear (for example figure \ref{fig:singularityF}). Figure %
\ref{fig:3Delectric} shows the full three-dimensional phase flow for a few
orbits with $\lambda >0$, which demonstrates the past-stability of $K_-$
and confirms the time-dependent stability of $F$.

\section{Dynamics of a magnetic field in axisymmetric Bianchi type I}
\label{ch:magnetic}

In the usual electromagnetism, Maxwell's equations treat electric and
magnetic fields in a symmetric manner. In the context of Bianchi
cosmologies, source-free pure electric and magnetic fields are
mathematically indistinguishable. The modification $\mathcal{L}_{H}$ breaks
this duality. Let us consider a homogeneous magnetic field along $x$-axis in the spacetime (\ref{eq:metric}). In parallel to the electric field of the previous section, we define $B$ as the magnetic field seen by a comoving observer:
\be
\mathbf{F}\equiv \frac{1}{2} F_{ab}\, dx^a \! \wedge \! dx^b = B(t) (e^{\alpha+\beta} dy)\wedge (e^{\alpha+\beta} dz).
\ee
Contrary to the electric case, equation (\ref{eq:vector}) is trivially satisfied for this Faraday tensor.  Instead the evolution of the comoving magnetic field is given by the Bianchi identity $\mathrm{d} \mathbf{F}=0$ which leads to 
\begin{equation}
\dot{B}=-2\left( H+\sigma \right) B.
\end{equation}%
In contrast to the peculiar dynamics of the electric field, there is no
modification to the evolution equation for the magnetic field since it comes
from the closedness of the field-strength 2-form. Then, we can solve
it easily to obtain 
\begin{equation}
B=\frac{B_{0}}{e^{2(\alpha +\beta )}}.  \label{eq:solB}
\end{equation}%
This simply means the magnetic field is adiabatically decaying due to the
expansion of the universe.

\begin{figure}[tbph]
\begin{center}
\includegraphics{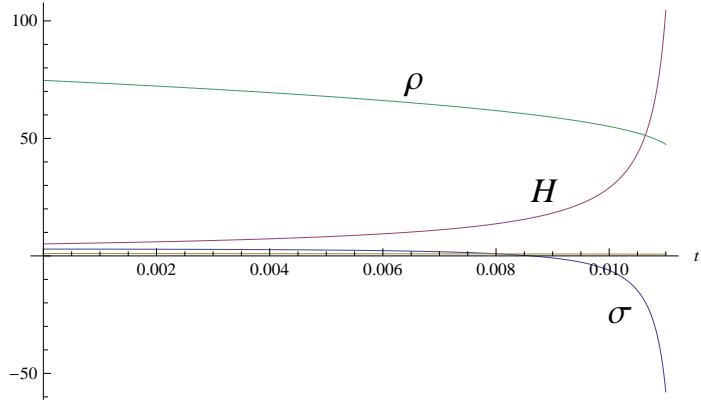}
\end{center}
\caption{Occurrence of the singularity for $\protect\lambda =-2,\protect%
\gamma =1$. Initial conditions are $H=5.1,\protect\sigma =2.9,B=1$. In fact,
the dynamics is more or less the same for any values of parameters or
initial conditions as long as $-\protect\lambda B^{2}>1$. In this case,
while the magnetic field is subdominant in the Friedmann equation (\protect
\ref{eq:FriedmannMag}), it still affects the evolution of $H$ and $\protect%
\sigma $ and causes their divergences. Note that $\protect\sigma %
/H\rightarrow -1$ as the singularity is approached. This implies the
approach towards the singularity appears as a flow into $K_{-}$ in the
expansion-normalised variables.}
\label{fig:singularity}
\end{figure}

The Einstein equations can be written in the following convenient from:
\begin{eqnarray}
H^{2} &=&\sigma ^{2}+\frac{1}{3}\rho +\frac{1}{6}B^{2}-\frac{2}{3}\lambda
\left( H+\sigma \right) \left( H-2\sigma \right) B^{2}, \label{eq:FriedmannMag} \\
\dot{H}+H^{2} &=&-\frac{1}{1+\lambda B^{2}}\left[ H^{2}+H\sigma +2\sigma
^{2}+\frac{2}{3}\left( \gamma -1\right) \rho -2\lambda H\left( H+2\sigma
\right) B^{2}\right]  \label{eq:raychaudhuri2} \\
&&\ + \frac{1}{2(1+\lambda B^{2})^{2}}\left[ \left( H+\sigma \right) ^{2}+%
\frac{\gamma -1}{3}\rho -\frac{1}{6}B^{2}-\frac{4}{3}\lambda \left( H+\sigma
\right) ^{2}B^{2}\right] ,  \notag \\
\dot{H}+\dot{\sigma} &=&-\frac{1}{1+\lambda B^{2}}\left[ 3\sigma \left(
H+\sigma \right) -3\lambda H\left( H+\sigma \right) B^{2}+\frac{\gamma }{2}%
\rho \right] .  \label{eq:key3}
\end{eqnarray}%
Again, there appears to be a problem for negative $\lambda $ when $\lambda
B^{2}\sim -1$. This time, the singularity stems from the Einstein equations
instead of the evolution equation for $F_{ab}$ as in the electric case. From
(\ref{eq:key3}), it is clear that $\dot{H}+\dot{\sigma}$ is positive
definite in an expanding universe ($H>0$) for initial conditions satisfying 
\begin{equation}
H+\sigma >0,\quad -\lambda B^{2}>1.  \label{eq:pathologicalM}
\end{equation}%
According to (\ref{eq:solB}), $B^2$ decays in this regime and consequently the system has to reach the singularity $\lambda B^2=-1$.
As one can see in figure \ref{fig:singularity}, the shear becomes negative
before reaching the singularity. The behaviour is insensitive to the initial
conditions or value of $\lambda $ whenever $-\lambda B^{2}>1$ initially.
This condition coincides with the one for the Horndeski modified term to
have a significant contribution to the dynamics. We conclude that $\lambda
<0 $ is pathological for the magnetic case too. 

While there is no divergent behaviour for positive $\lambda $, the energy
density of the magnetic field is not necessarily positive, in contrast to
the electric field case. This will not cause any problem for ordinary
expanding universes with $H+\sigma >0$ since the solution (\ref{eq:solB})
ensures monotonic decrease of $B$. However, it may result in unusual
behaviours when we go backwards in time. We repeat the dynamical system
analysis of the previous section and show that $\lambda >0$ is
cosmologically viable.

\subsection{Expansion-normalised autonomous system}

We introduce the following normalised variables: 
\begin{equation*}
\Sigma = \frac{\sigma }{H}, \ \ \ \ \ \Omega _M = \frac{B^2}{6H^2} ,\ \ \ \
\ \Omega _H = -\frac{2}{3}\lambda \left( 1+\Sigma \right) \left( 1-2\Sigma
\right) B^2 , \ \ \ \ \ \Omega _m = \frac{\rho }{3H^2} . 
\end{equation*}
The normalised Friedmann equation takes the canonical form 
\begin{equation}
1 = \Sigma ^2 + \Omega _H + \Omega _M + \Omega _m .  \label{eqM:Friedmann2}
\end{equation}
In contrast to the electric field, $\Omega _H$ is not positive definite
regardless of the sign of $\lambda $. Following the steps of the previous section, we rewrite (\ref%
{eq:raychaudhuri2}) as a defining equation for the deceleration parameter: 
\begin{eqnarray*}
q &=&-\frac{2\left( 1+\Sigma \right) \left( 1-2\Sigma \right) }{3\Omega
_{H}-2\left( 1+\Sigma \right) \left( 1-2\Sigma \right) }\left[ 1+\Sigma
+2\Sigma ^{2}+2(\gamma -1)\Omega _{m}+\frac{3\left( 1+2\Sigma \right) }{%
\left( 1+\Sigma \right) \left( 1-2\Sigma \right) }\Omega _{H}\right]  \\
&&-\frac{2\left( 1+\Sigma \right) ^{2}\left( 1-2\Sigma \right) ^{2}}{\left[
3\Omega _{H}-2\left( 1+\Sigma \right) \left( 1-2\Sigma \right) \right] ^{2}}%
\left[ \left( 1+\Sigma \right) ^{2}+(\gamma -1)\Omega _{m}-\Omega _{M}+\frac{%
2\left( 1+\Sigma \right) }{1-2\Sigma }\Omega _{H}\right] .
\end{eqnarray*}%
We derive the following evolution equations for the normalised variables: 
\begin{eqnarray}
\frac{d\Sigma }{d\alpha } &=&(1+q)(1+\Sigma )  \label{eqM:Sigma1} \\
&&+\frac{6\left( 1+\Sigma \right) \left( 1-2\Sigma \right) }{3\Omega
_{H}-2\left( 1+\Sigma \right) \left( 1-2\Sigma \right) }\left[ \Sigma \left(
1+\Sigma \right) +\frac{\gamma }{2}\Omega _{m}+\frac{3}{2\left( 1-2\Sigma
\right) }\Omega _{H}\right] ,  \notag \\
\frac{d\Omega _{M}}{d\alpha } &=&2\left( q-1-2\Sigma \right) \Omega _{M},
\label{eqM:Maxwell1} \\
\frac{d\Omega _{H}}{d\alpha } &=&-4(1+\Sigma )\Omega _{H}-\frac{1+4\Sigma }{%
\left( 1+\Sigma \right) \left( 1-2\Sigma \right) }\Omega _{H}\frac{d\Sigma }{%
d\alpha },  \label{eqM:Horndeski1} \\
\frac{d\Omega _{m}}{d\alpha } &=&\left( 2q-3\gamma +2\right) \Omega _{m}.
\label{eqM:matter}
\end{eqnarray}

\subsection{Fixed points and their stabilities}

We repeat the standard stability analysis. The structure of the state space
is analogous to the electric case. There are four fixed points residing in
the conventional magnetic Bianchi type I, and another with non-vanishing $%
\Omega _{H}$.

\begin{description}
\item[ Flat Friedmann universe : $F$ ] 
\begin{equation*}
\left( \Sigma ,\Omega _{M},\Omega _{H},\Omega _{m}\right) =\left(
0,0,0,1\right) 
\end{equation*}%
with eigenvalues 
\begin{equation}
(-4,-\frac{3}{2}\left( 2-\gamma \right) ,3\gamma -4).
\end{equation}%
Since the evolution equation of the magnetic field is well-behaved, the
linearisation can be carried out as usual. The stability does not change
from the usual magnetic cosmologies. $F$ is a future attractor for $\gamma
<4/3$. Note the zero eigenvalue in the radiation case ($\gamma =4/3$).

\item[ Magnetic Bianchi type I : $B$ ] 
\begin{equation*}
\left( \Sigma , \Omega _M , \Omega _H , \Omega _m \right) = \left( \frac{%
3\gamma -4}{4}, \frac{3}{16} \left( 2-\gamma \right) \left( 3\gamma -4
\right) ,0 , \frac{3}{8}(4 -\gamma ) \right) 
\end{equation*}
with eigenvalues 
\begin{equation}
\left(-\frac{3}{4}\left(2-\gamma \pm\sqrt{(2-\gamma )(3\gamma ^2 -17 \gamma
+18) }\right), -3\gamma \right).
\end{equation}
The existence condition is $\gamma > 4/3$ and it is always stable.

\item[ Kasner solutions : $K_{\pm }$ ] 
\begin{equation*}
\left( \Sigma , \Omega _M , \Omega _H , \Omega _m \right) = \left( \pm 1 ,
0, 0 ,0 \right) . 
\end{equation*}
The eigenvalues for $K_+$ are easily computed as 
\begin{equation}
(3(2-\gamma), -2, -8) ,
\end{equation}
indicating it is a saddle while being a future attractor for the subsystem $%
\Omega _m =0$. For $K_-$, we have the same indeterminacy of the linear
stability as was encountered in the electric case. It will be examined by
numerical analysis later.

\item[ $\boldsymbol{\Omega _H}$-dominated universe : $H_B$ ] 
\begin{equation*}
\left( \Sigma , \Omega _M , \Omega _H , \Omega _m \right) = \left( \frac{1}{2%
},0,\frac{3}{4},0 \right) . 
\end{equation*}
This fixed point does not itself represent a physical solution since it
corresponds to the limit $-\lambda(H+\sigma)(H-2\sigma) \rightarrow \infty$.
As in the electric case, this does not mean that it is dynamically
irrelevant since it can be reached asymptotically from finite $H$ and $%
\sigma $ (see figure \ref{fig:3Dmagnetic}). Note that $q=0$ for this fixed point which means that the back-reaction of the vector field exactly cancels out like for the Horndeski-dominated solution in the electric case.  Its eigenvalues are given by 
\begin{equation}
\left(-12, -3(\gamma +2 ), 6 \right).
\end{equation}
Therefore it is always a saddle point.
\end{description}

\subsection{Numerical analysis}

\begin{figure}[htbp]
\begin{center}
\includegraphics[width=0.9\textwidth]{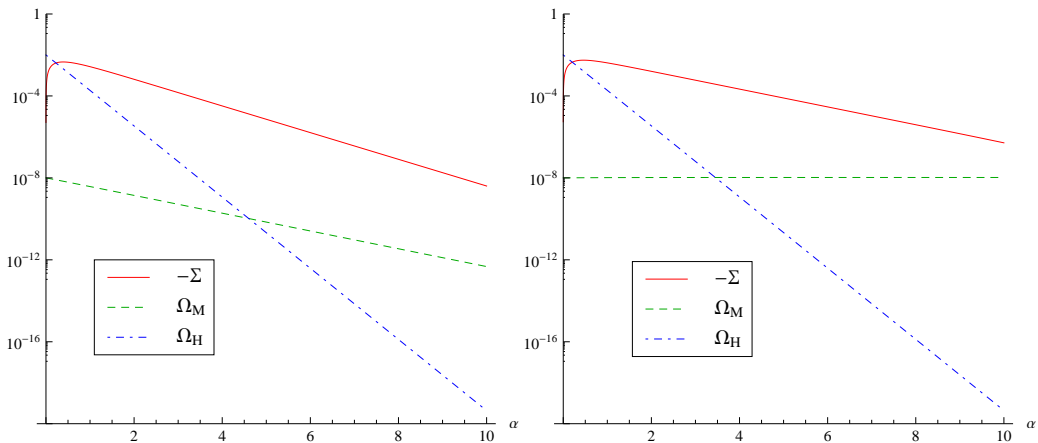}
\end{center}
\caption{Simulation of (\protect\ref{eqM:Sigma1})-(\protect\ref%
{eqM:Horndeski1}) close to the Friedmann solution $F$ for dust ($\protect\gamma=1$) to the left and radiation ($%
\protect\gamma=4/3$) to the right.  Initial conditions were $\Sigma= 10^{-8}$, $\Omega_M=10^{-8}$, $%
\Omega_H=10^{-2}$. }
\label{fig:CloseToFmagnetic}
\end{figure}

Since the uniform magnetic field decays adiabatically, the evolution close
to the Friedmann solution $F$ is trivial compared to the electric case. Near 
$F$, $\Omega _{M}$ evolves as in the conventional electromagnetism;
logarithmically decaying (constant at the linear level) in the radiation era
($\gamma =4/3$) because of the zero eigenvalue, and in proportion to $%
e^{-\alpha }$ in the dust era ($\gamma =1$), see \cite{Zel'dovich1970, Barrow1997, Barrow1998a}. It
follows immediately from its definition that $\Omega _{H}\propto
B^{2}\propto e^{-4\alpha }$, i.e., the evolution is independent of the
equation of state of the perfect fluid. See figure \ref{fig:CloseToFmagnetic}
for a simulation of the full non-linear equations (\ref{eqM:Sigma1})-(\ref%
{eqM:Horndeski1}) close to $F$.

\begin{figure}[htbp]
\begin{center}
\includegraphics[width=0.8\textwidth]{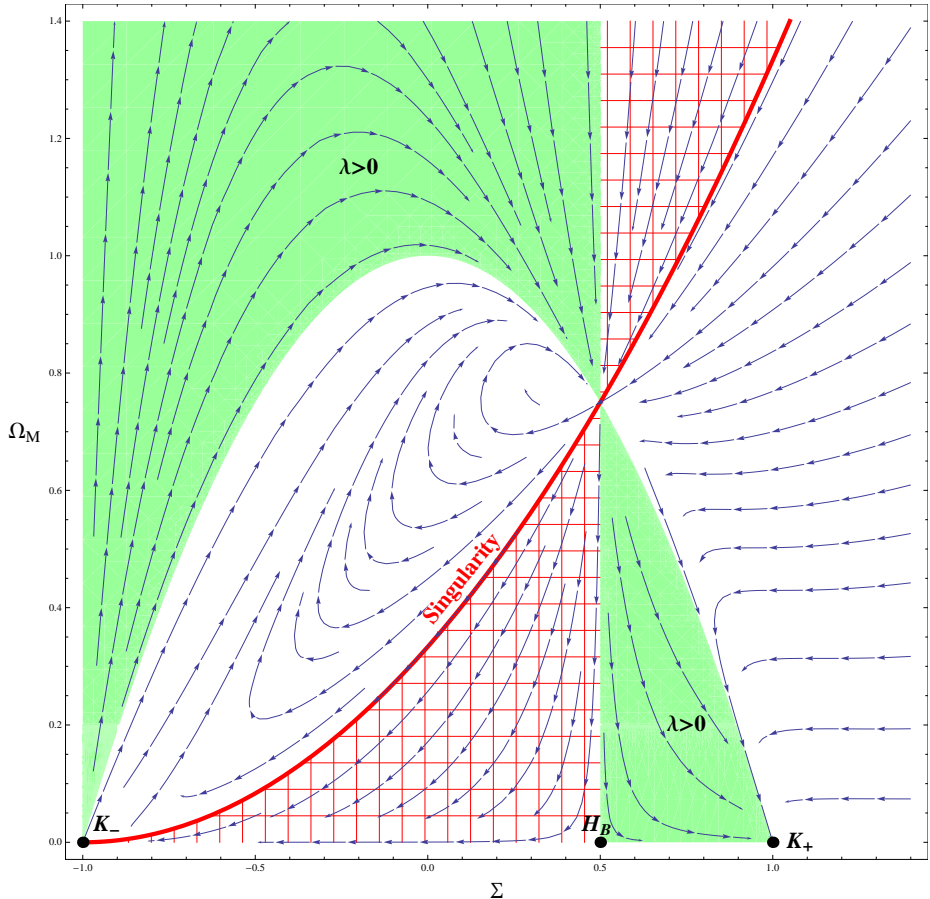}
\end{center}
\caption{Phase flow for the subsystem $\Omega_m=0$ in the magnetic case. The
green shaded region corresponds to $\protect\lambda>0$. The red bold line is
the position of the singularity, while the red mesh is the region where the
singularity is inevitable. In the meshed region, the flow goes towards $K_-$%
. For $\protect\lambda>0$, $K_-$ appears to be a past attractor.}
\label{fig:phaseM}
\end{figure}

\begin{figure}[htbp]
\begin{center}
\includegraphics[width=0.5\textwidth]{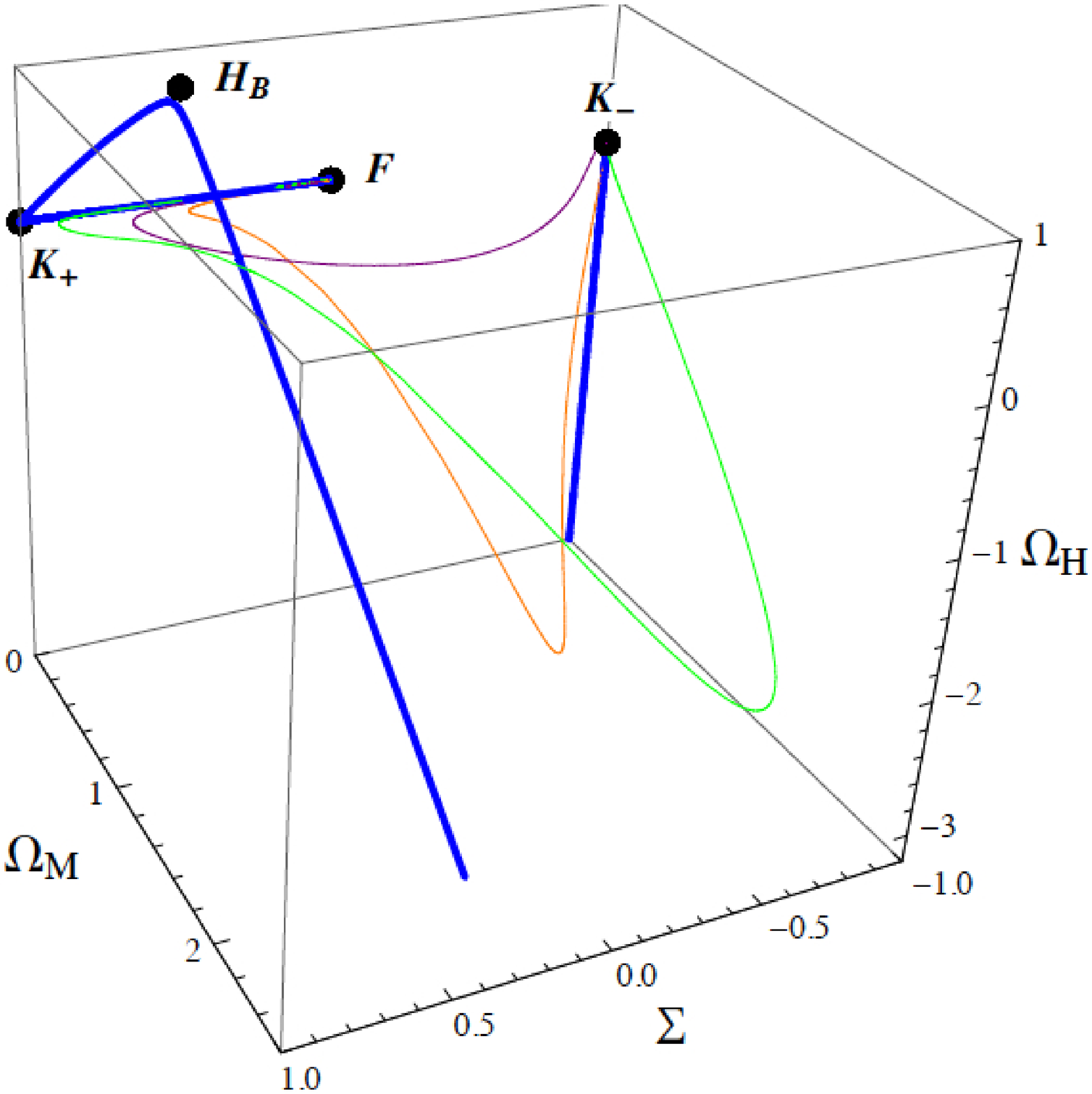}
\end{center}
\caption{ Simulations of the non-linear equations (\protect\ref{eqM:Sigma1}%
)-(\protect\ref{eqM:Horndeski1}) for a radiation fluid ($\protect\gamma=4/3$%
). All orbits correspond to a positive non-minimal coupling, $\protect\lambda%
>0$. For each orbit, phase flow is in the direction towards the Friedmann
solution F, and initial conditions are close to $K_-$ with non-vanishing
perfect fluid and the vector field, which underwrites the past stability of $%
K_-$. The thick blue orbit temporarily leaves the boundary of the box. After
it returns, it has transient periods close to $H_B$ and $K_+$ before
settling down at $F$.}
\label{fig:3Dmagnetic}
\end{figure}

It is again helpful to visualise the global structure of the phase space.
Figure \ref{fig:phaseM} shows a phase portrait of the invariant subset $%
\Omega _{m}=0$. The green shaded region corresponds to $\lambda >0$. It is
clearly seen that $K_{-}$ is a past attractor for $\lambda >0$ on this
subspace. Note the existence of stream lines directed towards $K_{-}$ in the
region $\lambda <0.$ They indicate orbit dependence of the stability for a
negative non-minimal coupling constant value. The red mesh corresponds to
the pathological region (\ref{eq:pathologicalM}) where a singularity, marked
with the red bold line, is inevitable. Note, from its definition, that $%
\Omega _{H}=0$ when $\Sigma =1/2$. Therefore, although it has no physical
significance, the point $(\Omega _{M},\Sigma )=(3/4,1/2)$ appears as if it
were a fixed point. Figure \ref{fig:3Dmagnetic} shows the full
three-dimensional phase flow for a few orbits with $\lambda >0$. Overall,
our simulations indicate that $K_{-}$ is a past attractor for $\lambda >0$.

\section{Observational constraints}
\label{ch:constraints}

In this section, we shall discuss the possibility of detecting any signature
of the theory with a positive coupling constant $\lambda $ in the standard
cosmology. Since the correction term comes with a factor of the space-time
curvature which is tiny in the units of Planck mass, one expects that a
large value of the parameter $\lambda $ is required to see any appreciable
effect. Combining the constraint on spatial anisotropy coming from big bang
nucleosynthesis \cite{Barrow1976} and the validity of the Maxwellian
electromagnetism in terrestrial environments, we will show that the growth
of the electric effect discussed in section \ref{ch:lineardynamics} must
have come to an end before the CMB temperature dropped below $T\sim 0.1$
MeV. After that time, $\Omega _{H}$ decays much faster than $\Omega _{M}$,
and so the effect of the Horndeski Lagrangian term is completely negligible
at late times. Since we restrict ourselves to the dynamics around FLRW in
this context, the magnetic case will not be discussed in detail due to its
trivial evolution seen in the previous section.

%This enables us to limit the cosmological scenarios to be discussed subsequently.  We shall see that the amplification of the normalized vector energy must come to an end before the start of Big Bang Nucleosynthesis.  Consequently the dynamics is unchanged at late times and we find a strong upper limit on the normalized Horndeski field energy today. 

%In this section we shall constrain some of the parameters and dynamical variables of our model observationally.  First we shall assume that the U(1) gauge field   represents the photons of electrodynamics.  We can then set a rough upper limit on the non-minimal coupling constant by demanding that modifications to conventional electrodynamics is negligible in earth-like environments. This enables us to limit the cosmological scenarios to be discussed subsequently.  We shall see that the amplification of the normalized vector energy must come to an end before the start of Big Bang Nucleosynthesis.  Consequently the dynamics is unchanged at late times and we find strong upper limits on the normalized Horndeski field energy today. To finish we also briefly discuss the situation when Horndeski's model is not viewed as a generalization of electrodynamics, but merely represents some vector-tensor theory, ie. the U(1) vector is not identified with photons.  This relaxes the constraint on the non-minimal coupling so that the amplification of the normalized vector energy may occur at late times, and we discuss cosmological and observational aspects of this scenario. 

A non-minimal coupling of electrodynamics to gravity is known to cause a
number of observational effects, such as frequency-dependent bending of
light in gravitational fields \cite{france94}. Since those effects must be
suppressed in terrestrial  environments, we require 
\begin{equation}
|\mathcal{L}_{H}|\ll \left\vert -\frac{1}{4}F^{2}\right\vert 
\label{constraintearth1}
\end{equation}%
evaluated around the Earth. In an orthonormal basis, the components of the
Riemann tensor for the Schwarzschild solution are of order $R_{abcd}\sim
R_{s}/r^{3}$, where $R_{s}$ is the Schwarzschild radius and $r$ is the
radial coordinate in Schwarzschild coordinates. Inserting the radius and
mass of the Earth, we find that the constraint (\ref{constraintearth1}) is
equivalent to
\begin{equation}
|\lambda |\ll \frac{r^{3}}{R_{s}}M_{pl}^{2}\sim 10^{90},
\label{constraintearth2}
\end{equation}%
in which the upper limit appears enormous since we have used a large mass
scale ($M_{pl}$) to normalise the non-minimal coupling constant (which is a
reasonable convention in the context of modified gravity). From the linear
analysis in section \ref{ch:lineardynamics}, we know that both $\Omega _{M}$
and $\Omega _{H}$ start to decay when 
\begin{equation}
\mathcal{R}\equiv \frac{\Omega _{H}}{\Omega _{M}}\simeq \lambda \frac{H^{2}}{%
M_{pl}^{2}}
\end{equation}%
is close to unity. At the time $\mathcal{R}\sim 1$, the expansion rate must
satisfy 
\begin{equation}
H\gg 10^{-45}M_{pl}
\end{equation}%
to be consistent with the constraint (\ref{constraintearth2}). Using the
approximation $T\sim (HM_{pl})^{\frac{1}{2}}$, where $T$ is the CMB
temperature, this imposes the following condition on the temperature at $%
\mathcal{R}\sim 1$: 
\begin{equation}
T_{\mathcal{R}\sim 1}\gg 10^{-23}M_{pl}\sim 0.1\;\text{MeV}.
\end{equation}%
Thus, to be consistent with the local constraint on terrestrial
electromagnetism, the growth of the normalised vector field energy is
required to end at latest around the time of big bang nucleosynthesis when $T_{%
\text{bbn}}= 0.1$-$1$ MeV. Since $\mathcal{R}$ is monotonically decreasing in
the perfect fluid dominated universe ($\mathcal{R}\propto e^{-4\alpha }$ in
the radiation-dominated epoch and $\mathcal{R}\propto e^{-3\alpha }$ in the
dust-dominated epoch), we set $\mathcal{R}\ll 1$ for $T\ll 1\ \text{MeV}$. It
follows that the energy density of the vector, both in the electric and the
magnetic case, will evolve as in the conventional electrodynamics, i.e. $%
E^{2},B^{2}\propto e^{-4\alpha }$ and therefore $\Omega _{H}\propto
e^{-4\alpha }$. Using that $T\propto e^{-\alpha }$, we obtain the relation 
\begin{equation}
\Omega _{H}^{\text{bbn}}\simeq \Omega _{H}^{0}\left( \frac{T_{\text{bbn}}}{%
T_{0}}\right) ^{4},
\end{equation}%
where $\Omega _{H}^{0}$ and $\Omega _{H}^{\text{bbn}}$ are the values of $%
\Omega _{H}$ today and at big bang nucleosynthesis, respectively. The CMB
temperature today is $T_{0}\simeq 10^{-4}\text{eV}$. We therefore have $%
\Omega _{H}^{\text{bbn}}\simeq 10^{40}\Omega _{H}^{0}$. The theory of big
bang nucleosynthesis requires that the universe was dominated by the CMB and
three effectively massless neutrino species at $T_{\text{bbn}}$, i.e. that $%
\Omega _{H}^{\text{bbn}}$ must be smaller than unity. This corresponds to
the following constraint at the present time: 
\begin{equation}
|\Omega _{H}^{0}|<10^{-40}.  \label{boundH}
\end{equation}%
We conclude that the effect of $\Omega _{H}$ on the evolution of the
universe becomes increasingly less significant after the big bang
nucleosynthesis. It is interesting to compare this constraint to that on a
uniform magnetic field in the conventional electrodynamics. In that case the
strongest limit on a uniform magnetic field today comes from the CMB
anisotropy and is \cite{barrow97} 
\begin{equation}
B_{0}\lesssim 10^{-9}\text{ Gauss}
\end{equation}%
or equivalently 
\begin{equation}
\Omega _{M}^{0}=\frac{B_{0}^{2}}{6M_{pl}H_{0}^{2}}\lesssim 10^{-12},
\label{boundM}
\end{equation}%
%
%\be
%\Omega_{M}^0 = \frac{B_0^2}{6\mu c^2 H_0^2} \lesssim 10^{-12}, 
%\label{boundM}
%\ee   
which is much weaker than the upper bound on the Horndeski part of the
Lagrangian (\ref{boundH}). Evaluating these constraints at decoupling time ($%
z\simeq 1100$) 
\begin{equation}
|\Omega _{H}^{\text{dc}}|\simeq \Omega _{H}^{0}1100^{4}\lesssim
10^{-28},\qquad \Omega _{M}^{\text{dc}}\simeq \Omega _{M}^{0}1100\lesssim
10^{-9},
\end{equation}%
we note that $|\Omega _{H}^{\text{dc}}|\ll \Omega _{M}^{\text{dc}}$.
Consequently the additional anisotropic stress induced by the non-minimal
coupling is completely negligible in the entire period between decoupling
time and today. Now, since the upper bound on a uniform magnetic field comes
from the modified temperature pattern in the CMB created by the shear (which
is sourced by the anisotropic stress), it is clear that the constraint so
far obtained is strong enough to exclude any signature of $\Omega _{H}$ in
the CMB, and (\ref{boundM}) holds also in Horndeski's generalised theory.
This result reflects the fact that $\Omega _{H}$ decays much faster than $%
\Omega _{M}$ after reaching $\mathcal{R}\sim 1$. In a similar manner, even a
direct detection of homogenous cosmological magnetic fields would not
improve the constraint on $\lambda $ either since the regime $\Omega
_{H}>\Omega _{M}$ must come prior to big bang nucleosynthesis.

%To summarize we have set a very strong upper limit on the present Horndeski contribution to the normalized vector energy (\ref{boundH}) by combining weak assumptions about Big Bang Nucleosynthesis and the size of the non-minimal coupling constant. Note that the upper limit on $\Omega_H$ corresponds to a constraint on the product of $\lambda$ and the Maxwell contribution to the vector field energy ($E^2$ or $B^2$).  Our considered cosmological scenario can therefore not improve the earth constraint on $\lambda$ (\ref{constraintearth2}).  In fact, even if a uniform magnetic field was detected around the upper limit $B\sim 10^{-9} \text{\,Gauss}$, the condition $\Omega_H<10^{-40}$ would correspond to $\lambda<10^{91}$ which already is much weaker than the earth constraint $\lambda\ll10^{91}$. We do expect, however, that cosmology can improve the earth limit on $\lambda$, but this requires going to the short wavelength limit which is beyond the scope of this paper. We have also explained why the upper limit on the magnetic field strength is unchanged in Horndeski's generalized theory. 

So far we have assumed that the vector field is identified with photons,
i.e. that the model is an extension of the ordinary electrodynamics. If this
is not the case, and the model is merely taken to be a hypothetical
vector-tensor theory, the bound on the non-minimal coupling constant (\ref%
{constraintearth2}) is not valid and the normalised vector field energy can
grow also at late times if $\lambda $ is enormous. To good approximation, we
have $\Sigma \gtrless 0$ when $\Omega _{M}\gtrless \Omega _{H}$, as noted in
section \ref{ch:lineardynamics}, and this could have some interesting
phenomenological consequences. Since the $\Delta T/T$ created in the CMB by
the shear is an integrated effect, which is proportional to \cite{thorsrud12}
\begin{equation}
\int_{\alpha _{\text{dc}}}^{\alpha _{0}}\Sigma d\alpha  \notag ,
\end{equation}%
a cancellation effect will occur if the peak value of the normalised vector
field energy (around $\mathcal{R}\sim 1$) occurs between decoupling time ($%
\alpha _{\text{dc}}$) and today ($\alpha _{0}$). First, one will have a
period with $\Omega _{H}>\Omega _{M}$ where both $\Omega _{M}$ and $\Omega
_{H}$ grow and $\Sigma <0$, followed by a period with $\Omega _{H}<\Omega
_{M}$ where both $\Omega _{M}$ and $\Omega _{H}$ decay and $\Sigma >0$; see
figure \ref{fig:CloseToDust}. By fine-tuning the non-minimal coupling
constant, the cancellation can be made exact leaving no effect on the CMB
despite having a relatively large amplitude of shear. Numerical simulations
show that this cancellation requires that $\Omega _{H}$ peaks at a low
redshift; as an example, $\Omega _{H}$ must peak at redshift $z\sim 0.6$ for
the initial conditions in figure \ref{fig:CloseToDust}. The shear still
needs to be consistent with the supernovae Ia data which is isotropic to the 
$1\%$ level, but this is significantly relaxed compared to the CMB bounds in
models where $\Sigma $ is positive or negative definite where $|\Sigma |$ is
never greater than $\sim 10^{-5}$. The possibility of such a cancellation
effect was mentioned in \cite{koivisto08,lim99} and to the best of our
knowledge, Horndeski's theory is the first instance of a Lagrangian for
which such a scenario may be dynamically realised.

%\begin{equation*}
%\left[ -\left( 2\dot{\alpha } +\dot{\beta } \right) \left( \dot{\alpha } +5\dot{\beta } \right) + \frac{1-6\lambda \left( \dot{\alpha }+\dot{\beta } \right) ^2 }{1+2\lambda \left( \dot{\alpha } +\dot{\beta } \right) ^2 } \left( \ddot{\alpha } +\ddot{\beta } \right) \right] 
% \end{equation*}

\section{Conclusion and outlook}
\label{ch:conclusion}

In this work, we have explored the cosmological consequences of a most
general vector-tensor theory that gives second-order field equations
equivalent to Maxwell's equations when evaluated in flat space-time. We have
focused on a simple dynamical setup which is well understood in the
conventional electromagnetism, namely the case of a uniform electric or
magnetic field in an axisymmetric Bianchi type I universe. We have
investigated the non-linear evolution equations by using several different
approaches, including the conventional dynamical systems analysis in terms
of Hubble normalised variables such as $\Omega _{M}$ and $\Omega _{H}$. In
contrast to the conventional electromagnetism, the theory treats electric
and magnetic fields in different ways and we have discussed each case
separately.

It has been found that the cases of a positive and negative non-minimal
coupling parameter $\lambda $ are drastically different. For $\lambda <0$,
we identified physical finite-time singularities where the deceleration
parameter diverges. In the electric case, the singularity comes from the
modified evolution equation for the electric field, while in the magnetic
case, it stems from the Einstein equations. In both cases, we identified a
range of initial conditions (which depend on $\lambda $) for which the
singularities are inevitable. Although we have been unable to show the
singularity is unavoidable during inflation, its mere existence in state
space is a problem in itself and suggests that the theory is pathological
for $\lambda <0$, regardless of the other matter fields in the universe. Our
result effectively rules out the possibility of generating anisotropy or
magnetic field during inflation by employing this type of non-minimal
coupling.

For $\lambda >0$, it appears that the modification is relatively harmless as
there is no singular behaviour and the Friedmann solution ($F)$ is the
future attractor for any perfect fluid satisfying $\gamma <4/3$. For
inflationary universes with $\gamma \ll 1$, we have found $F$ to be stable,
and consequently all anisotropies are washed out. In radiation and
dust-dominated universes, however, we find a rather interesting
phenomenology. Namely, in the electric case, the stability of $F$ is time
dependent and there is a transient period where both $\Omega _{M}$ and $%
\Omega _{H}$ grow. Both $\Omega _{M}$ and $\Omega _{H}$ reach their maximum
values around the time $\Omega _{M}\sim \Omega _{H}$, and then $\Omega _{M}$
decays adiabatically but the decrease of $\Omega _{H}$ is much faster.
Demanding the validity of the Maxwellian electromagnetism in terrestrial
environments, this peak value has to occur at the latest by the time of big
bang nucleosynthesis, which means the cosmological effect of $\Omega _{H}$
afterwards is always negligible. Therefore, all the cosmological constraints
after nucleosynthesis are based on the dynamics of $\Omega _{M}$ and hence
cannot be used to improve the upper bound on the non-minimal coupling
constant $\lambda $. Nevertheless, a huge amplification of electric field
energy may have taken place in the early universe ($\Omega _{M}\propto
e^{8\alpha }$ when $F$ is unstable). Even if electric fields are washed out
during inflation ($\Omega _{M}\propto e^{-4\alpha }$ if $\gamma \simeq 0$),
there could be enough time between reheating and big bang nucleosynthesis
for $\Omega _{M}$ to grow significantly if inflation lasts not much more
than $60$ e-folds and the reheating occurs at a relatively high energy
scale. After the amplification saturates, $\Omega _{H}$ decays quickly and
soon becomes negligible so that it does not cause any significant problems
at later times. We note that this amplification is characteristic of uniform
electric fields; for uniform magnetic fields, the dynamics near $F$ are
practically identical to the conventional electromagnetic case. It should also 
be pointed out that we have been ignoring any interaction between the electromagnetic
field and other matter species. In a realistic scenario, one expects the universe
becomes highly conducting at some stage and large scale electric fields quickly
decay away. To study this dissipation of electric fields due to the microphysics
of plasmas is beyond the scope of the present analysis for the homogeneous
fields.

Contrary to the minimally coupled Maxwell case, we found that the
generalised energy-momentum tensor may source both a positive and negative
shear. If the theory is merely taken to be some hypothetical vector-tensor
theory (not electromagnetism), so that the Hubble normalised field energy
may grow at late times, this opens up the possibility for cancellation
effects in the CMB. In particular we saw that if $\Omega _{H}$ peaks around
redshift $z\sim 0.6$, all CMB effects of the anisotropic expansion cancel
exactly. The possibility of such a cancellation effect was mentioned in \cite%
{koivisto08,lim99} and to the best of our knowledge, Horndeski's theory is
the first instance of a Lagrangian for which such a scenario may be
dynamically realised.

In this work we have neglected all spatial derivatives, and focused on
understanding a non-linear dynamics of the homogeneous fields. Physically, uniform
electric and magnetic fields correspond to a long wavelength limit of cosmic
electric or magnetic fields. This approach has been fruitful as it enabled
us to identify singularities at the background level and therefore rule out
any interesting cosmological application of this theory for a negative
non-minimal coupling. Although this makes the theory irrelevant in the
context of inflation, we have demonstrated that a non-trivial phenomenology
is possible in the early radiation-dominated universe. We point out that the
short-wavelength limit appears to be an interesting direction of further
investigation. Especially, when spatial derivatives are taken into account,
it would be interesting to see if an amplification of cosmic magnetic fields
could be possible before big bang nucleosynthesis. It is also necessary to
check if the model avoids instabilities from inhomogeneous perturbations.

\acknowledgments

We would like to thank Alan Coley, Gilles Esposito-Farese and Marco Peloso
for useful conversations. KY would like to thank David F. Mota and the
Institute of Theoretical Astrophysics in the University of Oslo for the
support and hospitality. MT would like to thank DAMTP at the University of Cambridge for the hospitality when this project was initiated.

\bibliographystyle{JHEP}
\bibliography{refs}

\providecommand{\href}[2]{#2}\begingroup\raggedright\begin{thebibliography}{10}

\bibitem{Clifton2011}
T.~Clifton, P.~G. Ferreira, A.~Padilla, and C.~Skordis, {\it {Modified Gravity
  and Cosmology}},  \href{http://xxx.lanl.gov/abs/1106.2476}{{\tt
  arXiv:1106.2476}}.

\bibitem{Milgrom2008}
M.~Milgrom and R.~H. Sanders, {\it {Rings and Shells of “Dark Matter” as
  MOND Artifacts}},  {\em The Astrophysical Journal} {\bf 678} (May, 2008)
  131--143, [\href{http://xxx.lanl.gov/abs/0709.2561}{{\tt arXiv:0709.2561}}].

\bibitem{Brans1961}
C.~Brans and R.~H. Dicke, {\it {Mach's Principle and a Relativistic Theory of
  Gravitation}},  {\em Physical Review} {\bf 124} (Nov., 1961) 925--935.

\bibitem{Nordtvedt1970}
K.~Nordtvedt, {\it {Post-Newtonian Metric for a General Class of Scalar-Tensor
  Gravitational Theories and Observational Consequences.}},  {\em The
  Astrophysical Journal} {\bf 161} (Sept., 1970) 1059.

\bibitem{Wagoner1970}
R.~V. Wagoner, {\it {Scalar-Tensor Theory and Gravitational Waves}},  {\em
  Physical Review D} {\bf 1} (June, 1970) 3209--3216.

\bibitem{Barrow1990}
J.~D. Barrow and K.-i. Maeda, {\it {Extended inflationary universes}},  {\em
  Nuclear Physics B} {\bf 341} (Sept., 1990) 294--308.

\bibitem{Ruzmaikina1971}
T.~V. Ruzmaikina and A.~A. Ruzmaikin, {\it {Gravitational Stability of an
  Expanding Universe in the Presence of a Magneric Field.}},  {\em Soviet
  Astronomy} {\bf 14} (June, 1971) 963.

\bibitem{Barrow1983}
J.~D. Barrow and A.~C. Ottewill, {\it {The stability of general relativistic
  cosmological theory}},  {\em Journal of Physics A: Mathematical and General}
  {\bf 16} (Aug., 1983) 2757--2776.

\bibitem{Barrow1988}
J.~D. Barrow and S.~Cotsakis, {\it {Inflation and the conformal structure of
  higher-order gravity theories}},  {\em Physics Letters B} {\bf 214} (Dec.,
  1988) 515--518.

\bibitem{Sotiriou2010}
T.~P. Sotiriou, {\it {f(R) theories of gravity}},  {\em Reviews of Modern
  Physics} {\bf 82} (Mar., 2010) 451--497,
  [\href{http://xxx.lanl.gov/abs/0805.1726}{{\tt arXiv:0805.1726}}].

\bibitem{Sotiriou2011}
T.~Sotiriou, B.~Li, and J.~Barrow, {\it {Generalizations of teleparallel
  gravity and local Lorentz symmetry}},  {\em Physical Review D} {\bf 83} (May,
  2011) 104030, [\href{http://xxx.lanl.gov/abs/1012.4039}{{\tt
  arXiv:1012.4039}}].

\bibitem{GOLDHABER1971}
A.~Goldhaber and M.~Nieto, {\it {Terrestrial and Extraterrestrial Limits on The
  Photon Mass}},  {\em Reviews of Modern Physics} {\bf 43} (July, 1971)
  277--296.

\bibitem{Barnes1979}
A.~Barnes, {\it {Cosmology of a charged universe}},  {\em The Astrophysical
  Journal} {\bf 227} (Jan., 1979) 1.

\bibitem{Barrow1984}
J.~D. Barrow and R.~R. Burman, {\it {Particle physics and cosmology: New light
  on heavy light}},  {\em Nature} {\bf 307} (Jan., 1984) 14--15.

\bibitem{Dolgov1981}
A.~Dolgov and Y.~Zeldovich, {\it {Cosmology and elementary particles}},  {\em
  Reviews of Modern Physics} {\bf 53} (Jan., 1981) 1--41.

\bibitem{Webb1999}
J.~Webb, V.~Flambaum, C.~Churchill, M.~Drinkwater, and J.~Barrow, {\it {Search
  for Time Variation of the Fine Structure Constant}},  {\em Physical Review
  Letters} {\bf 82} (Feb., 1999) 884--887,
  [\href{http://xxx.lanl.gov/abs/9803165}{{\tt 9803165}}].

\bibitem{Murphy2008}
M.~T. Murphy, J.~K. Webb, and V.~V. Flambaum, {\it {Revision of VLT/UVES
  constraints on a varying fine-structure constant}},  {\em Monthly Notices of
  the Royal Astronomical Society} {\bf 384} (Mar., 2008) 1053--1062,
  [\href{http://xxx.lanl.gov/abs/0612407}{{\tt 0612407}}].

\bibitem{Bekenstein1982}
J.~Bekenstein, {\it {Fine-structure constant: Is it really a constant?}},  {\em
  Physical Review D} {\bf 25} (Mar., 1982) 1527--1539.

\bibitem{Sandvik2002}
H.~v. Sandvik, J.~Barrow, and J.~a. Magueijo, {\it {A Simple Cosmology with a
  Varying Fine Structure Constant}},  {\em Physical Review Letters} {\bf 88}
  (Jan., 2002) 031302, [\href{http://xxx.lanl.gov/abs/0107512}{{\tt 0107512}}].

\bibitem{Barrow2002}
J.~D. Barrow and D.~F. Mota, {\it {Qualitative analysis of universes with
  varying alpha}},  {\em Classical and Quantum Gravity} {\bf 19} (Dec., 2002)
  6197--6212, [\href{http://xxx.lanl.gov/abs/0207012}{{\tt 0207012}}].

\bibitem{Barrow2002a}
J.~D. Barrow, J.~Magueijo, and H.~B. Sandvik, {\it {A cosmological tale of two
  varying constants}},  {\em Physics Letters B} {\bf 541} (Aug., 2002)
  201--210, [\href{http://xxx.lanl.gov/abs/0204357}{{\tt 0204357}}].

\bibitem{Barrow2002b}
J.~Barrow, H.~B. Sandvik, and J.~Magueijo, {\it {Behavior of varying-alpha
  cosmologies}},  {\em Physical Review D} {\bf 65} (Feb., 2002) 063504,
  [\href{http://xxx.lanl.gov/abs/0109414}{{\tt 0109414}}].

\bibitem{Uzan2003}
J.-P. Uzan, {\it {The fundamental constants and their variation: observational
  and theoretical status}},  {\em Reviews of Modern Physics} {\bf 75} (Apr.,
  2003) 403--455, [\href{http://xxx.lanl.gov/abs/0205340}{{\tt 0205340}}].

\bibitem{Barrow2012}
J.~Barrow and S.~Lip, {\it {Generalized theory of varying alpha}},  {\em
  Physical Review D} {\bf 85} (Jan., 2012) 023514,
  [\href{http://xxx.lanl.gov/abs/1110.3120}{{\tt arXiv:1110.3120}}].

\bibitem{Turner1988}
M.~Turner and L.~Widrow, {\it {Inflation-produced, large-scale magnetic
  fields}},  {\em Physical Review D} {\bf 37} (May, 1988) 2743--2754.

\bibitem{Ratra1992}
B.~Ratra, {\it {Cosmological 'seed' magnetic field from inflation}},  {\em The
  Astrophysical Journal} {\bf 391} (May, 1992) L1.

\bibitem{Calzetta1998}
E.~Calzetta, A.~Kandus, and F.~Mazzitelli, {\it {Primordial magnetic fields
  induced by cosmological particle creation}},  {\em Physical Review D} {\bf
  57} (June, 1998) 7139--7144, [\href{http://xxx.lanl.gov/abs/9707220}{{\tt
  9707220}}].

\bibitem{Giovannini2000}
M.~Giovannini, {\it {Magnetogenesis and the dynamics of internal dimensions}},
  {\em Physical Review D} {\bf 62} (Nov., 2000) 123505,
  [\href{http://xxx.lanl.gov/abs/0007163}{{\tt 0007163}}].

\bibitem{Lambiase2004}
G.~Lambiase and A.~Prasanna, {\it {Gauge invariant wave equations in curved
  space-times and primordial magnetic fields}},  {\em Physical Review D} {\bf
  70} (Sept., 2004) 063502, [\href{http://xxx.lanl.gov/abs/0407071}{{\tt
  0407071}}].

\bibitem{Kunze2005}
K.~E. Kunze, {\it {Primordial magnetic seed fields from extra dimensions}},
  {\em Physics Letters B} {\bf 623} (Sept., 2005) 1--9,
  [\href{http://xxx.lanl.gov/abs/0506212}{{\tt 0506212}}].

\bibitem{Bamba2007}
K.~Bamba and M.~Sasaki, {\it {Large-scale magnetic fields in the inflationary
  universe}},  {\em Journal of Cosmology and Astroparticle Physics} {\bf 2007}
  (Feb., 2007) 030--030, [\href{http://xxx.lanl.gov/abs/0611701}{{\tt
  0611701}}].

\bibitem{Kunze2008}
K.~Kunze, {\it {Primordial magnetic fields and nonlinear electrodynamics}},
  {\em Physical Review D} {\bf 77} (Jan., 2008) 023530,
  [\href{http://xxx.lanl.gov/abs/0710.2435}{{\tt arXiv:0710.2435}}].

\bibitem{Campanelli2008}
L.~Campanelli, P.~Cea, G.~Fogli, and L.~Tedesco, {\it {Inflation-produced
  magnetic fields in RnF2 and IF2 models}},  {\em Physical Review D} {\bf 77}
  (June, 2008) 123002, [\href{http://xxx.lanl.gov/abs/0802.2630}{{\tt
  arXiv:0802.2630}}].

\bibitem{Campanelli2008a}
L.~Campanelli, P.~Cea, G.~Fogli, and L.~Tedesco, {\it {Inflation-produced
  magnetic fields in nonlinear electrodynamics}},  {\em Physical Review D} {\bf
  77} (Feb., 2008) 043001, [\href{http://xxx.lanl.gov/abs/0710.2993}{{\tt
  arXiv:0710.2993}}].

\bibitem{Bamba2008a}
K.~Bamba, N.~Ohta, and S.~Tsujikawa, {\it {Generic estimates for magnetic
  fields generated during inflation including Dirac-Born-Infeld theories}},
  {\em Physical Review D} {\bf 78} (Aug., 2008) 043524,
  [\href{http://xxx.lanl.gov/abs/0805.3862}{{\tt arXiv:0805.3862}}].

\bibitem{Bamba2008b}
K.~Bamba, C.~Q. Geng, and S.~H. Ho, {\it {Large-scale magnetic fields from
  inflation due to Chern–Simons-like effective interaction}},  {\em Journal
  of Cosmology and Astroparticle Physics} {\bf 2008} (Nov., 2008) 013,
  [\href{http://xxx.lanl.gov/abs/0806.1856}{{\tt arXiv:0806.1856}}].

\bibitem{Campanelli2009}
L.~Campanelli and P.~Cea, {\it {Maxwell–Kosteleck\'{y} electromagnetism and
  cosmic magnetization}},  {\em Physics Letters B} {\bf 675} (May, 2009)
  155--158, [\href{http://xxx.lanl.gov/abs/0812.3745}{{\tt arXiv:0812.3745}}].

\bibitem{MosqueraCuesta2009}
H.~{Mosquera Cuesta} and G.~Lambiase, {\it {Primordial magnetic fields and
  gravitational baryogenesis in nonlinear electrodynamics}},  {\em Physical
  Review D} {\bf 80} (July, 2009) 023013,
  [\href{http://xxx.lanl.gov/abs/0907.3678}{{\tt arXiv:0907.3678}}].

\bibitem{Campanelli2009a}
L.~Campanelli, P.~Cea, and G.~Fogli, {\it {Lorentz symmetry violation and
  galactic magnetism}},  {\em Physics Letters B} {\bf 680} (Sept., 2009)
  125--128, [\href{http://xxx.lanl.gov/abs/0805.1851}{{\tt arXiv:0805.1851}}].

\bibitem{Kunze2010}
K.~E. Kunze, {\it {Large scale magnetic fields from gravitationally coupled
  electrodynamics}},  {\em Physical Review D} {\bf 81} (Feb., 2010) 043526,
  [\href{http://xxx.lanl.gov/abs/0911.1101}{{\tt arXiv:0911.1101}}].

\bibitem{Ford1989}
L.~Ford, {\it {Inflation driven by a vector field}},  {\em Physical Review D}
  {\bf 40} (Aug., 1989) 967--972.

\bibitem{Donnelly2010}
W.~Donnelly and T.~Jacobson, {\it {Coupling the inflaton to an expanding
  aether}},  {\em Physical Review D} {\bf 82} (Sept., 2010) 064032,
  [\href{http://xxx.lanl.gov/abs/1007.2594}{{\tt arXiv:1007.2594}}].

\bibitem{Gasperini1985}
M.~Gasperini, {\it {Inflation and broken Lorentz symmetry in the very early
  universe}},  {\em Physics Letters B} {\bf 163} (Nov., 1985) 84--86.

\bibitem{Carroll2004}
S.~Carroll and E.~Lim, {\it {Lorentz-violating vector fields slow the universe
  down}},  {\em Physical Review D} {\bf 70} (Dec., 2004) 123525,
  [\href{http://xxx.lanl.gov/abs/0407149}{{\tt 0407149}}].

\bibitem{Lim2005}
E.~Lim, {\it {Can we see Lorentz-violating vector fields in the CMB?}},  {\em
  Physical Review D} {\bf 71} (Mar., 2005) 063504,
  [\href{http://xxx.lanl.gov/abs/0407437}{{\tt 0407437}}].

\bibitem{Li2008}
B.~Li, D.~Mota, and J.~Barrow, {\it {Detecting a Lorentz-violating field in
  cosmology}},  {\em Physical Review D} {\bf 77} (Jan., 2008) 024032,
  [\href{http://xxx.lanl.gov/abs/0709.4581}{{\tt arXiv:0709.4581}}].

\bibitem{Zuntz2008}
J.~Zuntz, P.~Ferreira, and T.~Zlosnik, {\it {Constraining Lorentz Violation
  with Cosmology}},  {\em Physical Review Letters} {\bf 101} (Dec., 2008)
  261102, [\href{http://xxx.lanl.gov/abs/0808.1824}{{\tt arXiv:0808.1824}}].

\bibitem{Armendariz-Picon2010}
C.~Armendariz-Picon, N.~F. Sierra, and J.~Garriga, {\it {Primordial
  Perturbations in Einstein-Aether and BPSH Theories}},
  \href{http://xxx.lanl.gov/abs/1003.1283}{{\tt arXiv:1003.1283}}.

\bibitem{Zlosnik2008}
T.~Zlosnik, P.~Ferreira, and G.~Starkman, {\it {Growth of structure in theories
  with a dynamical preferred frame}},  {\em Physical Review D} {\bf 77} (Apr.,
  2008) 084010, [\href{http://xxx.lanl.gov/abs/0711.0520}{{\tt
  arXiv:0711.0520}}].

\bibitem{Meng2011}
X.-H. Meng and X.-L. Du, {\it {A Specific Case of Generalized Einstein-aether
  Theories}},  \href{http://xxx.lanl.gov/abs/1109.0823}{{\tt arXiv:1109.0823}}.

\bibitem{Nakashima2010}
M.~Nakashima and T.~Kobayashi, {\it {CMB Polarization in Einstein-Aether
  Theory}},  \href{http://xxx.lanl.gov/abs/1012.5348}{{\tt arXiv:1012.5348}}.

\bibitem{Koivisto2008a}
T.~Koivisto and D.~F. Mota, {\it {Vector field models of inflation and dark
  energy}},  {\em Journal of Cosmology and Astroparticle Physics} {\bf 2008}
  (Aug., 2008) 021, [\href{http://xxx.lanl.gov/abs/0805.4229}{{\tt
  arXiv:0805.4229}}].

\bibitem{Golovnev2008}
A.~Golovnev, V.~Mukhanov, and V.~Vanchurin, {\it {Vector Inflation}},  {\em
  Journal of Cosmology and Astroparticle Physics} {\bf 2008} (Feb., 2008) 6,
  [\href{http://xxx.lanl.gov/abs/0802.2068}{{\tt arXiv:0802.2068}}].

\bibitem{Bamba2008}
K.~Bamba and S.~D. Odintsov, {\it {Inflation and late-time cosmic acceleration
  in non-minimal Maxwell-\$F(R)\$ gravity and the generation of large-scale
  magnetic fields}},  {\em Journal of Cosmology and Astroparticle Physics} {\bf
  2008} (Jan., 2008) 20, [\href{http://xxx.lanl.gov/abs/0801.0954}{{\tt
  arXiv:0801.0954}}].

\bibitem{jimenez08}
J.~Beltr\'an~Jim\'enez and A.~L. Maroto, {\it {A cosmic vector for dark
  energy}},  {\em Phys.Rev.} {\bf D78} (2008) 063005,
  [\href{http://xxx.lanl.gov/abs/0801.1486}{{\tt arXiv:0801.1486}}].

\bibitem{himmetoglu08}
B.~Himmetoglu, C.~R. Contaldi, and M.~Peloso, {\it {Instability of anisotropic
  cosmological solutions supported by vector fields}},  {\em Phys.Rev.Lett.}
  {\bf 102} (2009) 111301, [\href{http://xxx.lanl.gov/abs/0809.2779}{{\tt
  arXiv:0809.2779}}].

\bibitem{carroll09}
S.~M. Carroll, T.~R. Dulaney, M.~I. Gresham, and H.~Tam, {\it {Instabilities in
  the Aether}},  {\em Phys.Rev.} {\bf D79} (2009) 065011,
  [\href{http://xxx.lanl.gov/abs/0812.1049}{{\tt arXiv:0812.1049}}].

\bibitem{himmetoglu08b}
B.~Himmetoglu, C.~R. Contaldi, and M.~Peloso, {\it {Instability of the ACW
  model, and problems with massive vectors during inflation}},  {\em Phys.Rev.}
  {\bf D79} (2009) 063517, [\href{http://xxx.lanl.gov/abs/0812.1231}{{\tt
  arXiv:0812.1231}}].

\bibitem{Koivisto2009}
T.~S. Koivisto, D.~F. Mota, and C.~Pitrou, {\it {Inflation from N-Forms and its
  stability}},  {\em Journal of High Energy Physics} {\bf 2009} (Mar., 2009)
  24, [\href{http://xxx.lanl.gov/abs/0903.4158}{{\tt arXiv:0903.4158}}].

\bibitem{Himmetoglu2009}
B.~Himmetoglu, C.~R. Contaldi, and M.~Peloso, {\it {Ghost instabilities of
  cosmological models with vector fields nonminimally coupled to the
  curvature}},  {\em Physical Review D} {\bf 80} (Sept., 2009) 44,
  [\href{http://xxx.lanl.gov/abs/0909.3524}{{\tt arXiv:0909.3524}}].

\bibitem{Golovnev2009}
A.~Golovnev, {\it {Linear perturbations in vector inflation and stability
  issues}},  {\em Physical Review D} {\bf 81} (Oct., 2009) 11,
  [\href{http://xxx.lanl.gov/abs/0910.0173}{{\tt arXiv:0910.0173}}].

\bibitem{Barrow1998}
J.~Barrow and J.~Levin, {\it {Chaos in the Einstein-Yang-Mills Equations}},
  {\em Physical Review Letters} {\bf 80} (Jan., 1998) 656--659,
  [\href{http://xxx.lanl.gov/abs/9706065}{{\tt 9706065}}].

\bibitem{Jin2005}
Y.~Jin and K.-i. Maeda, {\it {Chaos of Yang-Mills field in class A Bianchi
  spacetimes}},  {\em Physical Review D} {\bf 71} (Mar., 2005) 064007,
  [\href{http://xxx.lanl.gov/abs/0412060}{{\tt 0412060}}].

\bibitem{Barrow2005}
J.~D. Barrow, Y.~Jin, and K.-i. Maeda, {\it {Cosmological Co-evolution of
  Yang-Mills Fields and Perfect Fluids}},
  \href{http://xxx.lanl.gov/abs/0509097}{{\tt 0509097}}.

\bibitem{horndeski76}
G.~W. {Horndeski}, {\it {Conservation of Charge and the Einstein-Maxwell Field
  Equations}},  {\em J.Math.Phys.} {\bf 17} (1976) 1980--1987.

\bibitem{farese09}
G.~Esposito-Farese, C.~Pitrou, and J.-P. Uzan, {\it {Vector theories in
  cosmology}},  {\em Phys.Rev.} {\bf D81} (2010) 063519,
  [\href{http://xxx.lanl.gov/abs/0912.0481}{{\tt arXiv:0912.0481}}].

\bibitem{buchdahl79}
H.~A. Buchdahl, {\it {On a lagrangian for non-minimally coupled gravitational
  and electromagnetic fields}},  {\em J.Phys.} {\bf A12} (1979) 1037--1043.

\bibitem{Lovelock1971}
D.~Lovelock, {\it {The Einstein Tensor and Its Generalizations}},  {\em Journal
  of Mathematical Physics} {\bf 12} (Mar., 1971) 498.

\bibitem{Lovelock1972}
D.~Lovelock, {\it {The Four-Dimensionality of Space and the Einstein Tensor}},
  {\em Journal of Mathematical Physics} {\bf 13} (June, 1972) 874.

\bibitem{horndeski74}
G.~W. {Horndeski}, {\it {Second-Order Scalar-Tensor Field Equations in a
  Four-Dimensional Space}},  {\em International Journal of Theoretical Physics}
  {\bf 10} (Sept., 1974) 363--384.

\bibitem{scalartensor1}
T.~Kobayashi, M.~Yamaguchi, and J.~Yokoyama, {\it {Generalized G-inflation:
  Inflation with the most general second-order field equations}},  {\em
  Prog.Theor.Phys.} {\bf 126} (2011) 511--529,
  [\href{http://xxx.lanl.gov/abs/1105.5723}{{\tt arXiv:1105.5723}}].

\bibitem{scalartensor2}
A.~De~Felice, T.~Kobayashi, and S.~Tsujikawa, {\it {Effective gravitational
  couplings for cosmological perturbations in the most general scalar-tensor
  theories with second-order field equations}},  {\em Phys.Lett.} {\bf B706}
  (2011) 123--133, [\href{http://xxx.lanl.gov/abs/1108.4242}{{\tt
  arXiv:1108.4242}}].

\bibitem{scalartensor3}
A.~De~Felice and S.~Tsujikawa, {\it {Conditions for the cosmological viability
  of the most general scalar-tensor theories and their applications to extended
  Galileon dark energy models}},  {\em JCAP} {\bf 1202} (2012) 007,
  [\href{http://xxx.lanl.gov/abs/1110.3878}{{\tt arXiv:1110.3878}}].

\bibitem{scalartensor4}
C.~Charmousis, E.~J. Copeland, A.~Padilla, and P.~M. Saffin, {\it {Self-tuning
  and the derivation of a class of scalar-tensor theories}},  {\em Phys.Rev.}
  {\bf D85} (2012) 104040, [\href{http://xxx.lanl.gov/abs/1112.4866}{{\tt
  arXiv:1112.4866}}].

\bibitem{scalartensor5}
E.~J. Copeland, A.~Padilla, and P.~M. Saffin, {\it {The cosmology of the
  Fab-Four}},  \href{http://xxx.lanl.gov/abs/1208.3373}{{\tt arXiv:1208.3373}}.

\bibitem{scalartensor6}
S.~A. Appleby, A.~De~Felice, and E.~V. Linder, {\it {Fab 5: Noncanonical
  Kinetic Gravity, Self Tuning, and Cosmic Acceleration}},  {\em JCAP} {\bf
  1210} (2012) 060, [\href{http://xxx.lanl.gov/abs/1208.4163}{{\tt
  arXiv:1208.4163}}].

\bibitem{horndeski77}
G.~W. Horndeski and J.~Wainwright, {\it {Energy Momentum Tensor of the
  Electromagnetic Field}},  {\em Phys.Rev.} {\bf D16} (1977) 1691--1701.

\bibitem{blanc97}
V.~G. LeBlanc, {\it {Asymptotic states of magnetic Bianchi I cosmologies}},
  {\em Class.Quant.Grav.} {\bf 14} (1997) 2281--2301.

\bibitem{collins72}
C.~B. {Collins}, {\it {Qualitative magnetic cosmology}},  {\em Communications
  in Mathematical Physics} {\bf 27} (Mar., 1972) 37--43.

\bibitem{MTW}
C.~Misner, K.~Thorne, and J.~Wheeler, {\em Gravitation}.
\newblock W.H. Freeman, San Francisco, 1973.

\bibitem{Wainwright}
J.~Wainwright and G.~F.~R. Ellis, {\em {Dynamical Systems in Cosmology}}.
\newblock Cambridge University Press, 1997.

\bibitem{Zel'dovich1970}
Y.~B. Zel'dovich, {\it {The Hypothesis of Cosmological Magnetic
  Inhomogeneity.}},  {\em Soviet Astronomy} {\bf 13} (Feb., 1970) 608.

\bibitem{Barrow1997}
J.~Barrow, {\it {Cosmological limits on slightly skew stresses}},  {\em
  Physical Review D} {\bf 55} (June, 1997) 7451--7460,
  [\href{http://xxx.lanl.gov/abs/9701038}{{\tt 9701038}}].

\bibitem{Barrow1998a}
J.~Barrow and R.~Maartens, {\it {Anisotropic stresses in inhomogeneous
  universes}},  {\em Physical Review D} {\bf 59} (Dec., 1998) 043502,
  [\href{http://xxx.lanl.gov/abs/9808268}{{\tt 9808268}}].

\bibitem{Barrow1976}
J.~Barrow, {\it {Light elements and the isotropy of the Universe}},  {\em
  Monthly Notices of the Royal Astronomical Society} {\bf 175} (May, 1976)
  359--370.

\bibitem{france94}
R.~Lafrance and R.~C. Myers, {\it {Gravity's rainbow}},  {\em Phys.Rev.} {\bf
  D51} (1995) 2584--2590, [\href{http://xxx.lanl.gov/abs/hep-th/9411018}{{\tt
  hep-th/9411018}}].

\bibitem{barrow97}
J.~D. Barrow, P.~G. Ferreira, and J.~Silk, {\it {Constraints on a primordial
  magnetic field}},  {\em Phys.Rev.Lett.} {\bf 78} (1997) 3610--3613,
  [\href{http://xxx.lanl.gov/abs/astro-ph/9701063}{{\tt astro-ph/9701063}}].

\bibitem{thorsrud12}
M.~Thorsrud, D.~F. Mota, and S.~Hervik, {\it {Cosmology of a Scalar Field
  Coupled to Matter and an Isotropy-Violating Maxwell Field}},  {\em JHEP} {\bf
  1210} (2012) 066, [\href{http://xxx.lanl.gov/abs/1205.6261}{{\tt
  arXiv:1205.6261}}].

\bibitem{koivisto08}
T.~Koivisto and D.~F. Mota, {\it {Anisotropic Dark Energy: Dynamics of
  Background and Perturbations}},  {\em JCAP} {\bf 0806} (2008) 018,
  [\href{http://xxx.lanl.gov/abs/0801.3676}{{\tt arXiv:0801.3676}}].

\bibitem{lim99}
W.~Lim, U.~Nilsson, and J.~Wainwright, {\it {Anisotropic universes with
  isotropic cosmic microwave background radiation: Letter to the editor}},
  {\em Class.Quant.Grav.} {\bf 18} (2001) 5583--5590,
  [\href{http://xxx.lanl.gov/abs/gr-qc/9912001}{{\tt gr-qc/9912001}}].

\end{thebibliography}\endgroup

\end{document}